\newif\ifusenix
\newif\ifacm
\newif\ifmcom

\usenixfalse
\acmtrue
\mcomfalse

\ifusenix
	\documentclass[letterpaper,twocolumn,9pt]{article}
	\usepackage{usenix}
	\usepackage{times}
\fi
\ifacm
  \documentclass[9pt,sigconf]{acmart}
\fi

\ifmcom
  \documentclass{sig-alternate-10pt}
\fi

\usepackage{booktabs} 
\usepackage{pifont}
\usepackage{xcolor}
\usepackage{amsfonts}
\usepackage{balance}
\PassOptionsToPackage{hyphens}{url}
\usepackage{hyperref}
\usepackage{color}
\usepackage{graphics}
\usepackage{graphicx}
\usepackage{url}
\usepackage{listings}
\usepackage{multicol}
\usepackage{multirow}
\usepackage[scaled]{helvet}
\usepackage{rotating}
\usepackage{xspace}
\usepackage{algorithm}
\usepackage{comment}
\usepackage{enumitem}
\usepackage{amsmath}
\usepackage{mathrsfs}
\usepackage{cancel}
\usepackage{cleveref}
\usepackage{subfigure}
\usepackage{commath}
\usepackage[font=small,labelfont=bf]{caption}

\newcommand{\newtodo}[1]{\ClassWarning{NOT READY TO SUBMIT}{There is something left todo} \textcolor{blue}{}}

\newcommand{\highlightshep}[1]{{#1}}

\newcommand{\review}[1]{\ClassWarning{NOT READY TO SUBMIT}{There is something left todo} \textcolor{orange}{}}

\newcommand{\name}{mmReliable\xspace}
\newcommand{\constructivemultibeam}{constructive multi-beam\xspace}

\copyrightyear{2021}
\acmYear{2021}
\setcopyright{acmcopyright}\acmConference[SIGCOMM '21]{ACM SIGCOMM 2021 Conference}{August 23--27, 2021}{Virtual Event, USA}
\acmBooktitle{ACM SIGCOMM 2021 Conference (SIGCOMM '21), August 23--27, 2021, Virtual Event, USA}
\acmPrice{15.00}
\acmDOI{10.1145/3452296.3472924}
\acmISBN{978-1-4503-8383-7/21/08}



\definecolor{deepblue}{rgb}{0,0,0.5}
\definecolor{deepred}{rgb}{0.6,0,0}
\definecolor{deepgreen}{rgb}{0,0.5,0}
\definecolor{backcolour}{rgb}{0.95,0.95,0.92}

\def\beq{\begin{equation}}
\def\eeq{\end{equation}}
\def\beqa{\begin{eqnarray}}
\def\eeqa{\end{eqnarray}}
\def\beqan{\begin{eqnarray*}}
\def\eeqan{\end{eqnarray*}}

\def\R{{\mathbb{R}}}

\def\argmax{\mathop{\mathrm{arg\,max}}}

\def\x{\times}

\def\qed{\mbox{} \hfill $\Box$}
\def\tm1{t\! - \! 1}
\def\tp1{t\! + \! 1}

\newcommand{\abf}{\mathbf{a}}

\newcommand{\hbf}{\mathbf{h}}

\newcommand{\wbf}{\mathbf{w}}

\def\degree{^{\text{o}}}

\usepackage[explicit,noindentafter]{titlesec}
\titlespacing\section{1pt}{6pt plus 4pt minus 2pt}{0pt plus 2pt minus 2pt}
\titlespacing\subsection{1pt}{8pt plus 4pt minus 2pt}{1pt plus 2pt minus 2pt}
\titlespacing\subsubsection{1pt}{8pt plus 4pt minus 2pt}{1pt plus 2pt minus 2pt}
\begin{document}
\interfootnotelinepenalty=10000
\setlength{\belowdisplayskip}{2pt} \setlength{\belowdisplayshortskip}{2pt}
\setlength{\abovedisplayskip}{2pt} \setlength{\abovedisplayshortskip}{2pt}
\title{Two beams are better than one: Towards Reliable and High Throughput mmWave Links}


\author{Ish Kumar Jain, Raghav Subbaraman, Dinesh Bharadia}
\affiliation{
    \institution{University of California San Diego}
    \city{La Jolla}
    \state{CA}
    \country{USA}
    }
\renewcommand{\shortauthors}{IK Jain, R.Subbaraman, D.Bharadia}
\email{{ikjain, rsubbaraman, dineshb}@eng.ucsd.edu}

\renewcommand{\shorttitle}{Towards Reliable and High Throughput mmWave Links}

%

\ifacm

\begin{abstract}

Millimeter-wave communication with high throughput and high reliability is poised to be a gamechanger for V2X and VR applications. However, mmWave links are notorious for low reliability since they suffer from frequent outages due to blockage and user mobility. We build \name, a reliable mmWave system that implements multi-beamforming and user tracking to handle environmental vulnerabilities. It creates constructive multi-beam patterns and optimizes their angle, phase, and amplitude to maximize the signal strength at the receiver. Multi-beam links are reliable since they are resilient to occasional blockages of few constituent beams compared to a single-beam system. We implement \name on a 28 GHz testbed with 400 MHz bandwidth, and a 64 element phased array supporting 5G NR waveforms. Rigorous indoor and outdoor experiments demonstrate that \name achieves close to 100\% reliability providing 2.3x improvement in the throughput-reliability product than single-beam systems.\footnote{This is an extended version of a paper published in Sigcomm'21}

\end{abstract}

\fi

\maketitle

\ifusenix
    
\fi

\ifmcom
    
\fi

\section{Introduction}\label{sec:intro}

5G New Radio (NR) is expected to support cutting-edge applications such as vehicular (V2X), factory automation, autonomous driving, and remote surgery~\cite{5gnr, wang2020demystifying}. A key requirement for such applications beyond the high data rate is exceptionally high reliability, defined as the fraction of time during which the link does not suffer an outage~\cite{ji2018ultra,jain2019impact}. 
As such, 5G NR utilizes millimeter-wave (mmWave, FR2) frequencies due to abundant bandwidth (400MHz-2GHz) that can provide the high-data-rate compared to traditional sub-6 GHz FR1 (20-100MHz bandwidth) bands 
~\cite{Rappaport2013Millimeter}. However, unlike its sub-6 GHz counterparts, mmWave uses directional beams to reduce the impact of higher path loss at mmWave frequencies and is highly susceptible to blockage and user movement. For instance, a measurement study of 28 GHz network deployment in 
Chicago~\cite{narayanan2019first} shows that blockage by humans always causes link outage, which
is detrimental to applications requiring high-reliability~\cite{ji2018ultra}.

A large fraction of existing work focuses on improving mmWave link establishment~\cite{ jeong2015random,barati2016initial, sur201560, abari2016millimeter,hassanieh2018fast}, throughput~\cite{sur2016beamspy, zhou2017beam,palacios2018adaptive,Xarray2020} and coverage~\cite{zhou2019robot,wei2017facilitating}; while the link reliability requirement takes the backseat. The prevailing methods aim to reduce link outages by performing \textit{beam-training}, a process where the mmWave base station and client tediously scan multiple beams to establish a directional link. Reducing beam-training time does not prevent outages as the process is usually reactive~\cite{zhu2014demystifying, jeong2015random,barati2015directional,zhou2012efficient,barati2016initial, sur201560, abari2016millimeter,hassanieh2018fast} and kicks in only after a significant degradation in link quality. To improve reliability, some authors propose proactive approaches that constantly track the client using side-channel information from out-of-band LTE/WiFi~\cite{sur2017wifi,nitsche2015steering} or external sensors like GPS,  radar or lidar~\cite{haider2018listeer, wei2017pose,va2015beam, woodford2021spacebeam}. These solutions are not self-reliant, have limited accuracy, and possibly require sensitive information like user location~\cite{wei2017pose, va2015beam}; making the system dependant on external information and difficult to deploy with existing standards. In contrast, we aim to create the ideal mmWave link: one that can be created and maintained with high reliability (no link outage), supports high data rates, and is easy to integrate into current standards.


%

\begin{figure}[t!]
    \centering
    \includegraphics[width=0.35\textwidth]{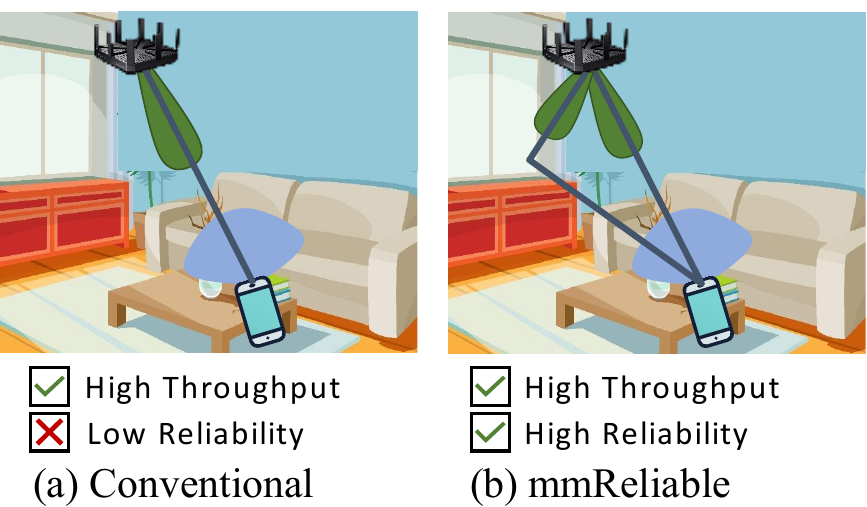}
    \caption{\name utilizes beams with multiple lobes (\textit{Multi Beam}) to provide high reliability compared to a single-lobe beams.}
    \label{fig:prototype}
\end{figure}


In this paper, we introduce \name, a system that achieves both reliable and high-data-rate mmWave links while being protocol compliant. Inspired by sub-6 GHz communication, \name delivers on reliability and throughput by exploiting the multipath diversity in environmental reflections. These reflections are typically strong for mmWave and can sustain the link even if the direct path is occluded or unavailable to establish an independent link. We observe that conventional single-beam links which utilize the reflectors are still unreliable because they are prone to outage due to blockage or misalignment. \name uses single radio frequency (RF) chain and phased array to create custom beam patterns with multiple directional beams (\textit{multi-beam}) that are each aligned towards direct and reflected paths as shown in Fig.~\ref{fig:prototype}(b). 
The idea here is that the probability of multiple beams simultaneously facing a blockage is low, reducing the possibility of link outage and increasing reliability. 


An obvious implication of splitting a single beam into a multi-beam is reduced power radiated per beam to conserve the Total Radiated Power (TRP) for FCC compliance~\cite{fcc}. The reduction in per-beam power could reduce the received signal strength and lead to throughput loss at the receiver, i.e., trading off throughput for reliability. It turns out we can break this trade-off: we observe that the received signal is the sum of multiple copies of the transmitted signal, one from every beam in a multi-beam as shown in Fig.~\ref{fig:prototype}(b). If we ensure constructive addition of these copies at the receiver (Section~\ref{sec:creatingmultibeam}), the received power and throughput can be increased to a value higher than that of a single beam while keeping TRP constant.

Let's take an intuitive example of how multi-beam improves throughput. Consider a mmWave channel consisting of two paths, and two beams are aligned along those two-channel paths. For simplicity, assume the path loss is unity for each path, and the signal traveled along each path incurs the same phase. \name's transmitter splits the total power of $a^2$ equally into two beams with power $\frac{a^2}{2}$ on each beam, i.e., amplitude $\frac{a}{\sqrt{2}}$ on each beam, which propagates through the channel and is received at the Omni-receiver. Since the phase align for both paths, the received signal adds in signal domain, i.e. total amplitude of the received signal $\frac{a}{\sqrt{2}} + \frac{a}{\sqrt{2}} = \sqrt{2}a$. The received signal upon converting to SNR (signal-to-noise ratio) is proportional to $2a^2$.
In contrast, a traditional mmWave single-beam link does not exploit the second path and would provide an SNR proportional to $a^2$, which is 3 dB lower compared to the multi-beam case. Higher SNR results in lower bit error rates and allows higher modulation and coding schemes, leading to higher throughput for multi-beam. In Section~\ref{sec:creatingmultibeam}, we go beyond this simple example to consider practical channels and show that, if the beam directions, phase, and power per beam of multi-beam are chosen optimally, then multi-beam will always achieve SNR higher than single beam \textit{even in wireless channels with weak multipath, while using the exact total radiated power}. A multi-beam created using optimal parameters is hereafter referred to as \constructivemultibeam.

As a first, we develop an algorithm to establish a \constructivemultibeam link with low-overhead and protocol compliance. To achieve \constructivemultibeam, one needs to estimate the optimal beam directions, beam-phase, and power per beam. Our key insight here is to decouple this problem into two: finding beam directions and estimating the per beam power and phase.
\name first learns the directions of strong paths in the environment (i.e., paths viable for communication) during the mandatory beam-training phase. This could be done using exhaustive beam-scanning or any other improved algorithm~\cite{jeong2015random,barati2016initial,ghasempour2018multi, jog2019many}. Then, with just two additional channel probes per viable beam, it learns the optimal power and phase to be applied to each beam in the multi-beam. Only two-to-three viable beams exist in typical environments due to the sparse nature of reflection clusters at mmWave~\cite{zhu2014demystifying, telecom2019analysis}; therefore, \name's algorithmic overhead remain fixed and independent of the number of elements in the antenna array.

Once the \constructivemultibeam is established, the challenge is in continuously maintaining it, as the wireless channel changes due to blockages and mobility. Link blockages could completely occlude one or more beams of a multi-beam. Mobility adversely affects the established links by causing misalignment between transmit and receive beams, degrading both throughput and reliability. Multi-beam management design has to address some uniquely challenging aspects. In contrast to a single-beam system where only one beam is misaligned, in a multi-beam system, all the beams get misaligned simultaneously due to mobility. In addition, using a phased array system with a single RF chain means that signals from multiple beam-directions are always superposed into one. The superposition makes it difficult to assess the effects of blockage and misalignment.

We develop a beam-maintenance algorithm, which proactively re-aligns each of the beam parameters (direction, phase, and amplitude) in a multi-beam and continuously maintains its constructive nature. First, \name uses a super-resolution algorithm to tease apart the properties of individual beams from their sum. Next, it uses the estimated per beam properties to measure power loss and infer the underlying cause, i.e., blockage or mobility. Instead of waiting for beams to degrade, \name proactively makes this inference and optimizes per-beam properties, and maintains \constructivemultibeam. \name uses low-overhead standard-compliant channel probes that are embedded in the communication waveform to achieve the elusive goal of creating and maintaining reliable high throughput mmWave links, free from sporadic outages.

We build a first-of-its-kind software-defined 5G NR testbed on 28 GHz mmWave bands with 64 elements phased array that supports channels of up to 400 MHz~\cite{jain2020mmobile}. We implement \name using 5G NR compliant reference signals. We perform experiments at 28 GHz in our university area by deploying \name in various indoor and outdoor scenarios with link distances of up to 80 m. Through empirical measurements, we find that indoor walls and large outdoor buildings are powerful reflectors; the median attenuation of the reflected path with respect to the direct path in outdoor scenarios is only 5 dB. Our evaluation shows that in scenarios where user movement and blockage co-occur, \name has reliability close to 100\% while maintaining an average throughput of 1.5 bits/sec/Hz as compared to the single-beam reactive baseline, which has a reliability of 65\% and on-average throughput of 1 bits/sec/Hz. The throughput-reliability product is improved by 2.3$\times$ compared to the best reactive baseline. The artifacts for \name are available online\footnote{Artifacts link: \url{https://wcsng.ucsd.edu/mmreliable}}.



\section{Background and Motivation}\label{sec:motivation}



Directional beamforming in mmWave creates the need for beam-management schemes for link creation and maintenance. Beam-management can be divided into two broad functions: the first is \textbf{beam-training}, and involves the search and creation of a beam~\cite{zhu2014demystifying, jeong2015random,barati2015directional,zhou2012efficient,barati2016initial, sur201560, abari2016millimeter,hassanieh2018fast}. The second is \textbf{beam-maintenance} and involves the upkeep of the beam despite environmental factors such as blockages, mobility, and fading. Even if multi-beam is resilient to blockages, it is not practical unless it can be efficiently created and maintained under the constraints of existing standards. The efficacy of the beam-maintenance function is especially critical to enabling reliable mmWave links with low overhead. Protocols such as 5G-NR and IEEE 802.11ad contain exclusive signaling and control provisions for such beam-management~\cite{5gnr,ieee80211ad,giordani2018tutorial}.

\vspace{-2mm}
\subsection{\name: Making Multi-beam Practical}
While single-beams have been the default choice for communication, there are some recent interest in using multi-beams to enable auxiliary mmWave capabilities. Multi-beams have been shown to aid with the beam-training process~\cite{hassanieh2018fast, aykin2019multi,aykin2019smartlink,feng2017dealing}. In~\cite{hassanieh2018fast}, the authors employ multi-beams for fast beam-training, but use them to set up single-beam links for communication. In~\cite{aykin2019multi}, the authors propose a beam-training method to create multi-beams for blockage resilience. However, they do not derive or discuss that a multi-beam link is better than a single-beam link for throughput. In addition, all of~\cite{hassanieh2018fast, aykin2019multi,aykin2019smartlink,feng2017dealing} depend on repeated reactive beam-training to re-establish the link when it is blocked or misaligned, leading to intermittent outages. In contrast, we propose multi-beam as the new standard for beamforming in mmWave communication systems. We show that multi-beams are reliable while providing superior throughput to single-beams in all scenarios. As a first, we develop an efficient algorithm to estimate parameters and create multi-beams in state-of-the-art phased arrays. Our algorithm is protocol compatible and can be deployed over any beam-training scheme. 


\vspace{-2mm}
\subsection{Beam Maintenance for Reliable mmWave}\label{sec:motivation_beam_maintenance}
\begin{figure}
    \centering
    \includegraphics[width=0.45\textwidth]{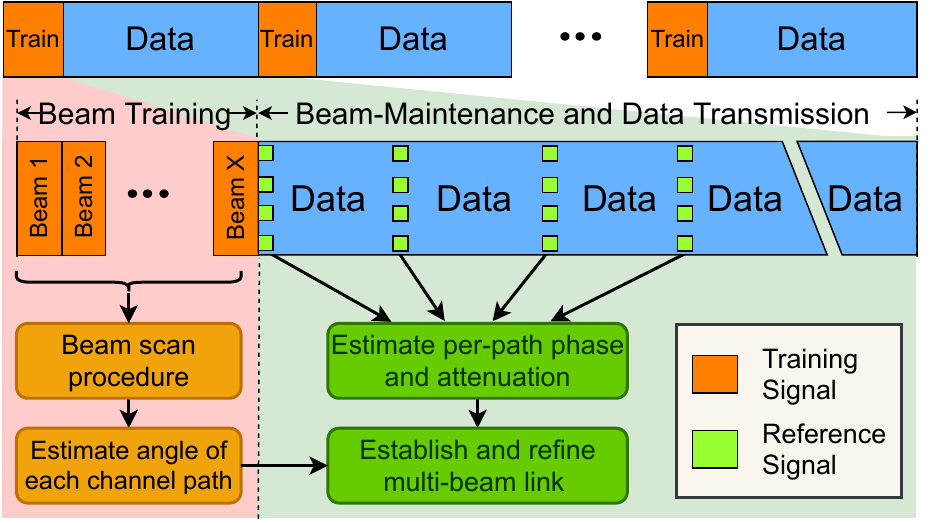}
    \caption{\highlightshep{Building blocks of \name beam management that leverages initial beam training for angle estimation and reference signaling for multi-beam establishment and maintenance.}}
    \label{fig:5GNR_mmReliable_ovreview}
\end{figure}

Beam maintenance is important to reduce the frequency and overhead of repeated beam-training as shown in Fig.~\ref{fig:5GNR_mmReliable_ovreview}. For instance, in the context of 5G-NR, a beam-training phase could take up to 5 ms to probe 64 beam directions with a default periodicity of 20 ms, leading to a 25\% overhead~\cite{giordani2018tutorial}. Even if a more efficient training algorithm from~\cite{hassanieh2018fast, aykin2019multi,aykin2019smartlink,feng2017dealing,ganji2021unblock} is used, the overhead will make beam-management solely based on beam-training intractable. The high overhead of beam-training can be avoided by carefully utilizing intrinsic features of mmWave protocols. They use known reference signals interspersed with data communication to perform beam-maintenance and refinement (See Fig. \ref{fig:5GNR_mmReliable_ovreview}). These reference signals provide channel measurements with an option to set any desired beam. Due to the sparsity of these reference signals, they cannot be used for a full-fledged beam-training, and only minor changes to the beamforming vector can be made using few channel estimates. A low-overhead beam-maintenance scheme is, therefore, a must for reliable mmWave links.


While beam-maintenance for single beam links have been studied by some authors~\cite{aykin2020mamba,marzi2016compressive,jain2020mmobile}, there is no work that addresses multi-beam links. Due to this gap, the utility of multi-beam for communication is lessened despite its benefits. We are the first to propose a low overhead beam-maintenance scheme using multi-beam. Our approach periodically updates the per-beam angle, phase, and amplitude with negligible overhead, requiring only three channel estimates for a 2-beam multi-beam and five estimates for a 3-beam multi-beam. The entire process is completed within 0.6 ms for the latter, making multi-beam maintenance possible with low overhead. With this, our work makes a compelling case for multi-beam and addresses the end-to-end challenges in their creation and maintenance.

\section{Design of Multi-beam system}\label{sec:creatingmultibeam}
This section describes how \name creates and leverages multi-beam for reliable, high-throughput mmWave communication. 
We first show how links can be made reliable to blockage by using multiple beams at once. Then, we derive mathematically optimal beamforming weights to maximize the signal-to-noise ratio (SNR) at the receiver. We show that optimal beamforming is nothing but \constructivemultibeam for most typical mmWave channels. 


We make an interesting observation that \constructivemultibeam is channel-dependent and depends on the number of paths, their angles, relative phase, and attenuation. Naturally, to implement \constructivemultibeam, we need to estimate these parameters. We propose a two-step procedure to estimate these parameters using the beam-management framework shown in Fig.~\ref{fig:5GNR_mmReliable_ovreview}. We leverage any standard beam-training procedure to calculate the angle of each path~\cite{ghasempour2018multi,jog2019many}. Then, we design a low-overhead method using reference signals to estimate the per-path phase and attenuation. Finally, we use these estimates to set up and maintain the optimal \constructivemultibeam despite the adversity of blockage and user mobility.

\vspace{-2mm}
\subsection{Multi-beam links are more reliable}
To compare various techniques on equal footing, we formally define link reliability as the fraction of time when the link is available for communication within a large observation interval. Link outage (due to natural effects) or procedures like beam-training reduce the reliability as they temporarily render the link unavailable for communication. Therefore, we can express reliability as:
\begin{equation}
    \text{Reliability} = 1-\text{Prob}(\text{Outage})
\end{equation}
We empirically compute the probability of outage as the fraction of duration where the SNR is below a minimum threshold. The directional nature of single-beam mmWave links makes them susceptible to link outages since blockage \& user mobility effects can reduce SNR by up to 30 dB~\cite{maccartney2017rapid} (Fig. \ref{fig:snr_blockage_timeseries}). Multi-beams prevent these outages by avoiding a single point of failure.

Multi-beam links are reliable even under the impact of blockage events. Consider a blockage probability $\beta \;(0\le\beta\le1)$, which represents the fraction of time the link is in outage during the observation interval. For simplicity, we assume the beams in our multi-beam are blocked independently. Thus, the probability that $k$ beams simultaneously experience an outage is $\beta^k$. By definition, the reliability would be $1-\beta$ for the single-beam case and $1-\beta^k, k\ge2$ for the multi-beam case. Multi-beam provides higher reliability because it prevents the link from suffering an outage due to blockage, unlike its single-beam counterparts~\cite{sur2016beamspy,maccartney2017rapid}. In practice, blockages may be correlated, and scenarios could arise where two or more paths are blocked simultaneously. While no solution can prevent link outage if all paths are blocked (e.g., the user inadvertently blocks all antennas with their body), multi-beam links can be sustained as long as there is at least one unblocked path to the receiver. Simultaneous blockage of multiple paths is always less probable than blockage of a single path~\cite{sur2016beamspy}; therefore, multi-beam links are \textit{always} more reliable than single-beam links.
\begin{figure*}
    \centering
    \includegraphics[width=0.48\textwidth]{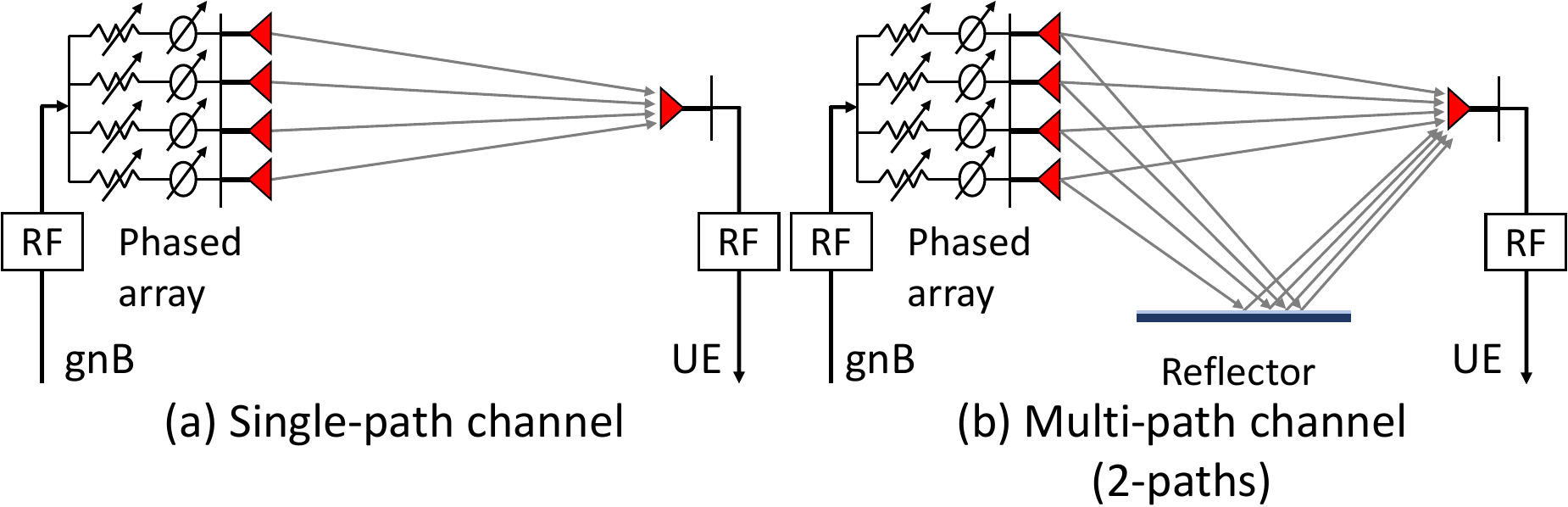}
    \includegraphics[width=0.46\textwidth]{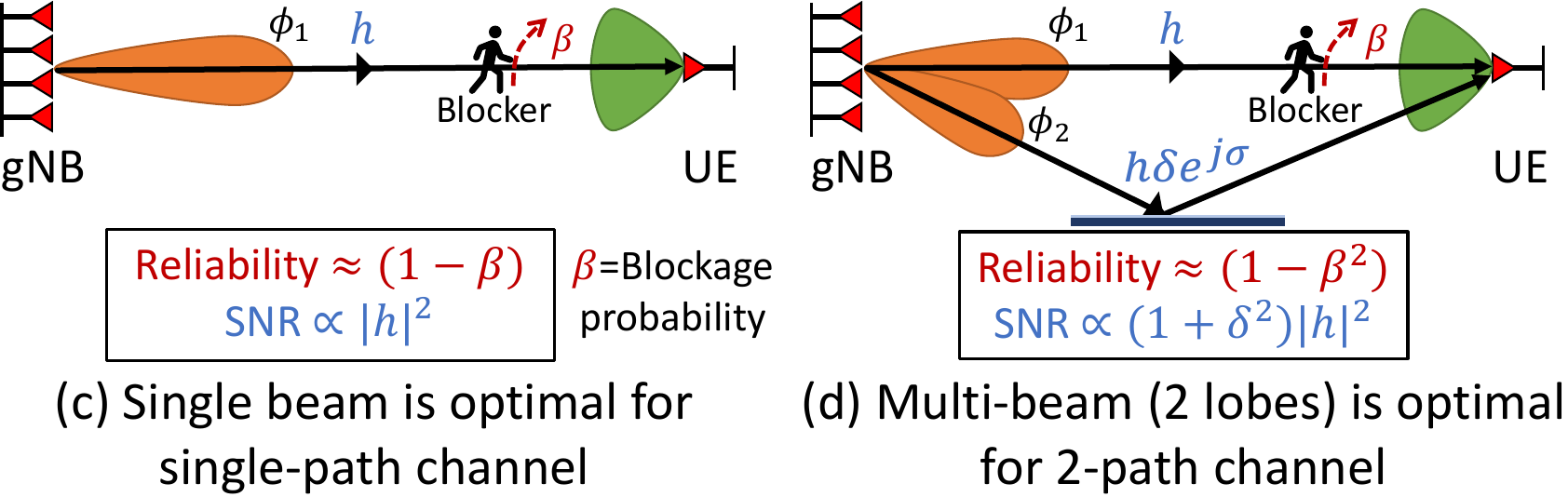}
    \caption{Constructive multi-beam enhances the throughput \& reliability of mmWave link. (a) System with a single-path channel, (b) With multipath (2-path) channel. (c) The optimal beamforming technique for a single-path channel is a single beam with all the energy directed along that channel path. (d) A constructive multi-beam with two lobes is more reliable and provides higher SNR compared to a single beam.}
    \vspace{2pt}
    \label{fig:reliability_snr}
\end{figure*}

\vspace{-2mm}
\subsection{Multi-beam Throughput Model}
While multi-beams are intuitively resilient to blockage by virtue of having multiple beams separated in angle, each beam's power is reduced since the total power is split amongst multiple beams. One may wonder whether this leads to reduced throughput since it could mean less power is incident on the receiver. But this is not the case, and it can be shown that a multi-beam link with optimal constructive combining and power-control can, in fact, provide higher throughput than a single-beam link. To prove this, we introduce a model for a multi-antenna base station transmitting to the user device using a mmWave link and derive the optimal beamforming solution that maximizes the received SNR. We make two important observations. First, we show that a single-beam system is optimal only when the wireless channel consists of a single path from the transmitter to the receiver. A single-beam system is sub-optimal for a multipath channel, which is generally the case even for mmWave links~\cite{sun2013multi}. We then show that a multi-beam with two beams is optimal for a two-path channel and maximizes the received SNR. We conclude with a note on generalizing our formulation to an arbitrary number of beams in multi-beam.

\noindent $\blacksquare\;$ \textbf{Primer on optimal beamforming:}
We derive an expression for optimal beamforming weights at a multi-antenna base-station (gNB phased array) communicating with a single antenna receiver using mmWave (Multi-antenna receiver discussed in Section~\ref{sec:multibeamUE}). The gNB uses a uniform linear phased array with $N$ antenna elements. Beamforming is implemented by applying a $N\times 1$ beam weights vector $\wbf$ at each antenna using a combination of phase shifters and attenuators (Fig.~\ref{fig:reliability_snr}(a)). Our goal is to determine the optimal $\wbf$ that maximizes the received SNR. 

Let $s$ be the transmit signal from the gNB, the received signal $y$ is expressed as:
\begin{equation}
    y = \hbf^T \wbf s + \eta,
\end{equation}
where $\hbf$ is  $N\times 1$ channel from $N$ transmit antennas to one receive antenna and $\eta$ is white Gaussian noise at the receiver. The capacity or maximum throughput of a wireless link is evaluated in term of SNR (the ratio of signal power to noise power) as: 
\begin{equation}\label{eq:snr}
    \text{SNR} = {||\hbf^T \wbf ||^2 P_s}/{P_\eta},
\end{equation}
 where $P_s$ is the average transmit power (without transmit array gain) and $P_\eta$ is the noise power\footnote{To derive the expression for SNR: $E[yy^H] = E[(\hbf^H \wbf s + n)(\hbf^H \wbf s + n)^H]] = \|\hbf^H \wbf\|^2 E[ss^H] + E[nn^H]$ and we denote $P_s = E[ss^H]$ and $P_\eta =E[nn^H]$.}.
 
 Our goal is to estimate beamforming weights $\wbf$, that maximizes the SNR. Using Cauchy-Schwartz inequality, it follows that $||\hbf^T \wbf ||$ is maximized when $\hbf^{*}$ (complex conjugate of $\hbf$) and $\wbf$ align in vector space~\cite{tse2005fundamentals}. Therefore, \textit{the optimal weight vector is channel-dependent}. Intuitively, the weight vector cancels the phases in the channel and creates an inner-product form that maximizes the absolute value of their product. Additionally, the weights need to be unitary to keep the power constant, i.e., $\|\wbf\|=1$. This way, we get the optimal weights ${\wbf}^{\text{opt}}$ as:
\begin{equation}\label{eq:wmrc}
    {\wbf}^{\text{opt}} = {\hbf^*}/{||\hbf||},
\end{equation}
The optimal beamforming provides the highest SNR of $\frac{||\hbf||^2P_s}{P_\eta}$. The practical wireless channel---even mmWave channel---consists of multiple paths (direct or reflected paths).
Since the optimal beamforming vector depends on the channel $\hbf$, it is affected by the nature of these paths.

\noindent $\blacksquare\;$ \textbf{Single-beam is optimal for a single-path channel:} Extending our discussion, it is natural to wonder how to provide a structure to the general nature of optimal beamforming vectors (${\wbf}^{\text{opt}}$). Consider the simplest of wireless channels with a single path from the transmitter to the receiver. It turns out that the single beam used by conventional mmWave systems is optimal for such a channel. To see this, we represent the single-path channel vector by $\hbf^{\text{single}}$ and derive the optimal weights. A single-path channel is entirely defined by two parameters: the direction of departure $\phi_1$ and complex attenuation $h$:
\begin{equation}\label{eq:1pathchannel}
    \hbf^{\text{single}}[n] = h e^{-j2\pi \frac{d}{\lambda} (n-1) \sin(\phi_1)},
\end{equation}
where $n$ is the transmit-antenna index, $d$ is the antenna spacing, and $\lambda$ is the wavelength of carrier frequency ($d=\frac{\lambda}{2}$ in our phased array). Using (\ref{eq:wmrc}) and (\ref{eq:1pathchannel}), we obtain the optimal weights for this channel as $\wbf_{\phi_1}=\frac{ \hbf^{\text{single}^*}}{\| \hbf^{\text{single}}\|} $ which can be simplified to: 
\begin{equation}\label{eq:wphi1}
\begin{split}
     \wbf_{\phi_1}=\frac{1}{\sqrt{N}}[1, e^{j2\pi \frac{d}{\lambda}\sin(\phi_1)},\ldots,  e^{j2\pi(N-1) \frac{d}{\lambda}\sin(\phi_1)}]^T,
\end{split}
\end{equation}
which is the familiar single-beam weight vector~\cite{sur2016beamspy, zhou2017beam, sur201560,zhou2018following}.


\noindent $\blacksquare\;$ \textbf{Constructive multi-beam is optimal for multipath channel:}
Since the optimal beam is channel dependent, let us observe how it changes when we introduce a reflector into the channel as shown in Fig.~\ref{fig:reliability_snr}(b). We continue to use $h$ for the complex attenuation of the first path. The second path's attenuation is expressed as $h \delta e^{j\sigma}$, where $\delta \in \R^+$ and $\sigma \in [0,2\pi]$ are respectively the relative attenuation and phase shift of the second path with respect to the first. Now, we can write the expression of two-path channel as:
\begin{equation}\label{eq:hmultieq}
    \hbf^{\text{multi}}[n] = h e^{-j2\pi \frac{d}{\lambda} (n-1) \sin(\phi_1)} + h \delta e^{j\sigma} e^{-j2\pi \frac{d}{\lambda} (n-1) \sin(\phi_2)}
\end{equation}
where $\phi_1$ and $\phi_2$ are the respective directions of departure of the two paths. 
If we naively use the single-beam weights from (\ref{eq:wphi1}) here, it ignores the second path and severely attenuates power along $\phi_2$ (Fig.~\ref{fig:reliability_snr}(c)) resulting in an approximate SNR of:
\begin{equation}
    \text{SNR}^{\text{single}} = {||(\hbf^{\text{single}})^T \wbf_{\phi_1} ||^2 P_s}/{P_\eta} \approx {|h|^2P_s}/{P_\eta}.
\end{equation}
In contrast, the optimal weights for the multipath channel are obtained from (\ref{eq:wmrc}) as $\textbf{w}^{\text{multi}} = {\hbf^{\text{multi}^*}}/{\|\hbf^{\text{multi}}\|}$ which is visualized in Fig. \ref{fig:reliability_snr}(d). From Eq.~(\ref{eq:snr}), the SNR of the optimal weights $\textbf{w}^{\text{multi}}$ is:
\begin{equation}
    \text{SNR}^{\text{multi}} = \|h^{\text{multi}}\|^2 P_s/P_\eta \approx {(1+\delta^2)|h|^2P_s}/{P_\eta}.
\end{equation}
In comparison to the naive single-beam approach, the optimal SNR is higher by a factor of $1+\delta^2$, which provides 2x gain (3dB higher SNR) for two equally strong paths ($\delta=1$) as we have seen in Section~\ref{sec:intro}. In general, $\textbf{w}^{\text{multi}}$ is the optimal \constructivemultibeam weights that split the transmit signal along both the paths such that they maximally combine at the receiver. We discuss in Appendix~\ref{appendix:A} how this result can be generalized to show that a k-multi-beam is the optimal solution for a general k-path channel.

\begin{figure} [!t]
    \subfigure{{\includegraphics[width=0.45\columnwidth]{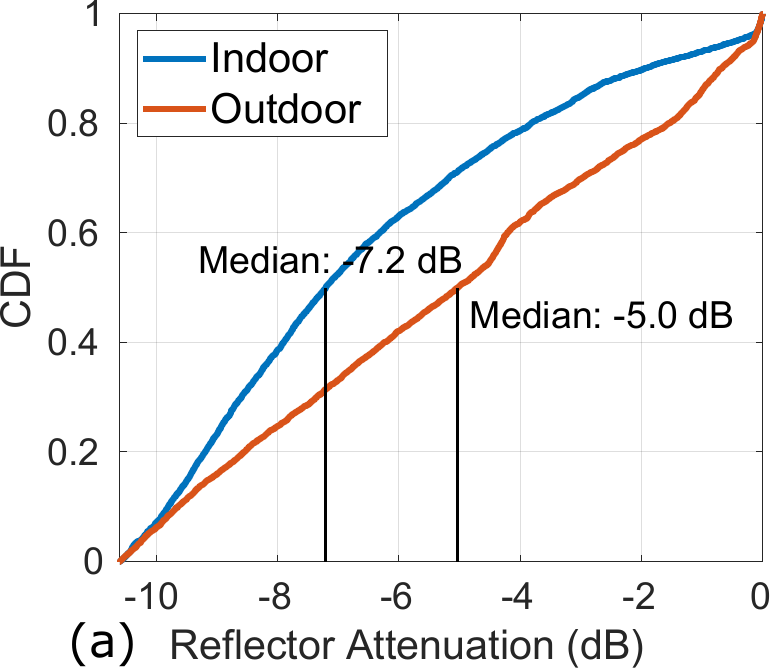}\label{fig:reflector_relative_power_cdf}}}
    \subfigure{{\includegraphics[width=0.53\columnwidth]{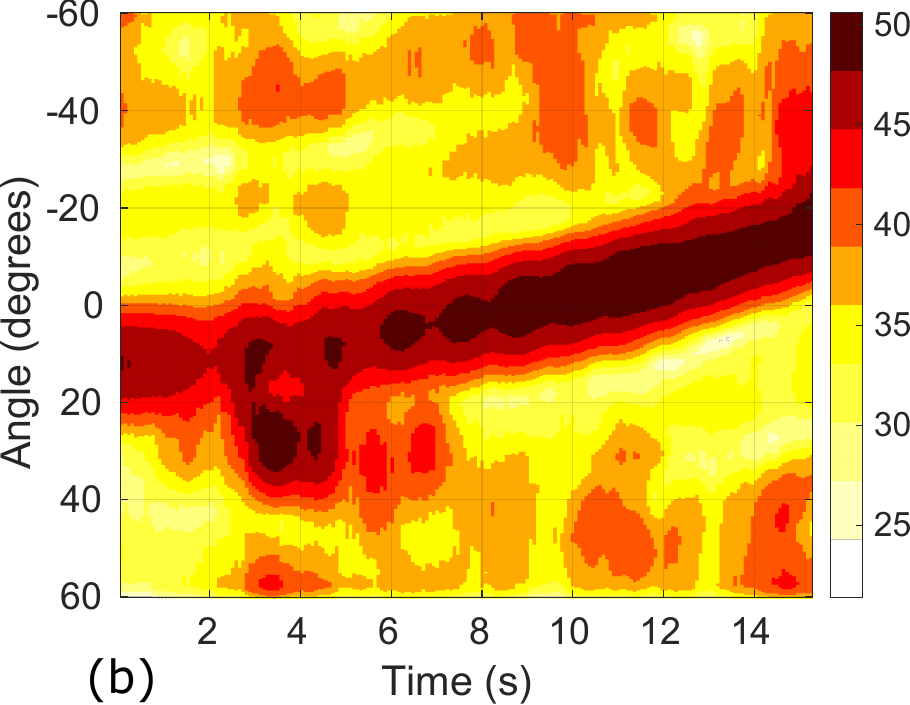}\label{fig:reflector_movt_example_noblock}}}
    \caption{(a) CDF of the relative attenuation of the strongest reflected path compared to the direct path at over various locations. (b) A heatmap of strong paths in the environment while the UE moves, strong reflectors appear at different points in time.}
    \label{fig:reflector_characterize}
\end{figure}

\noindent $\blacksquare\;$ \textbf{Strength of mmWave multipath:} At this point, one may ask if multiple paths exist in practical mmWave channels, and if so, how strong are they? There are numerous measurement studies that indicate the presence of strong multipath at both 28 GHz and 60 GHz~\cite{Rappaport2013Millimeter,zhu2014demystifying, telecom2019analysis,nitsche2014ieee,xu2002spatial,bas2019outdoor}. These studies show that in a typical deployment, reflectors like metals, concrete walls, and tinted glass strongly reflect signals while attenuating them by about 5-6 dB ($\delta\approx0.5$). In addition, we perform a measurement study of reflector strengths at various indoor (5m-10m link) and outdoor (10m-80m link) locations (Overall 10K data points). At each location, we perform a full 120$\degree$ scan to identify the distribution of signal strength across space. Shown in Fig.~\ref{fig:reflector_relative_power_cdf}, common reflectors cause 1-10 dB attenuation (relative to the direct path) with a median attenuation of 7.2 dB indoors and 5 dB outdoors. Fig.~\ref{fig:reflector_movt_example_noblock} shows a particular data point where signal strengths along all angles are evaluated during user motion. Sometimes, reflectors as strong as the direct path exist and can be utilized to establish a multi-beam link. 

\vspace{-2mm}
\subsection{Creating constructive multi-beam}\label{sec:creating_constructive_multibeam}
\label{subsec:creating_mb}
So far, we have established that \constructivemultibeam is the optimal beamforming technique for a typical mmWave channel. To achieve optimal beamforming, we need to estimate the channel $\hbf[n]$ at each antenna at the gNB, the complexity of which scales with the total number of antennas. In sub-6 GHz communication, the channel at each antenna is readily measured as each antenna could be connected to a separate RF chain. In contrast, mmWave hardware usually consists of a single RF chain attached to a large phased array with many antennas. Performing channel estimations for each antenna element is intractable in practical mmWave systems.

The intuition behind \constructivemultibeam is that it optimally uses the available multipath. We make the key observation that in mmWave, the number of multipath reflections is usually sparse~\cite{zhu2014demystifying, telecom2019analysis}; therefore, there must be a simpler way to measure $h[n]$. Indeed, from (\ref{eq:hmultieq}), we see that the channel can be linearly decomposed into its constituent single-beam directions. Said differently: we can reconstruct $h[n]$ using channel measurements on individual multipath directions in the environment. Once we have $h[n]$, the optimal weights follow using (\ref{eq:wmrc}).

Consider our example from the previous sub-section, where two paths were present in the environment at angles $\phi_1,\phi_2$. If the relative attenuation $\delta$ and the relative phase $\sigma$ between each path is measured, we can create the \constructivemultibeam weights by simple addition:
\begin{equation}\label{eq:beampattern}
    \textbf{w}(\phi_1,\phi_2,\delta,\sigma) = \frac{(\textbf{w}_{\phi_1}+{\delta} e^{-j{\sigma}}\textbf{w}_{\phi_2})}{\|(\textbf{w}_{\phi_1}+{\delta} e^{-j{\sigma}}\textbf{w}_{\phi_2})\|},
 \end{equation}
The denominator ensures that the TRP is conserved by enforcing $\|\textbf{w}\|=1$. We then appropriately quantize the beamforming vector phases \& amplitudes to be compatible with our phased array. In~\cite{aykin2019multi}, a sub-optimal multi-beam is created by splitting the array into sub-arrays, each responsible for a particular component beam. In contrast, \name utilizes phase and amplitude control to create the optimal channel-dependent multi-beam.

\noindent $\blacksquare\;$ \textbf{Estimating parameters for \constructivemultibeam:}
Reliably estimating the parameters to establish a multi-beam is integral to maintaining its constructive nature. In the two-beam case, we need to estimate $\delta$ (relative attenuation, $\delta \in \R^+$) and $\sigma$ (relative phase shift, $0<\sigma\le 2\pi$) of the reflected path w.r.t. the direct path. We already know the directions $\phi_1,\phi_2$ of the two paths from the beam-training phase. We denote the narrow-band complex wireless channel along each path as $h_1$ and $h_2$, respectively.
If the gNB measures both $h_1$ and $h_2$, then it can estimate the relative amplitude and phase shift using their ratio. A simple method would be to measure each beam's channel one-by-one by setting the beam to $\phi_1$ and then $\phi_2$, as suggested in~\cite{aykin2019multi}. However, hardware offsets such as Carrier Frequency Offset (CFO) and Sampling Frequency Offset (SFO) cause time-varying and sometimes unpredictable channel phases, making these channel estimates unreliable~\cite{palacios2018adaptive, dunna2020scattermimo, kotaru2015spotfi}. The channel magnitude is the one thing that remains fixed despite these offsets. Therefore, we develop an estimation method using channel magnitude alone.
\begin{figure}[t]
    \centering
    \includegraphics[width=0.45\textwidth]{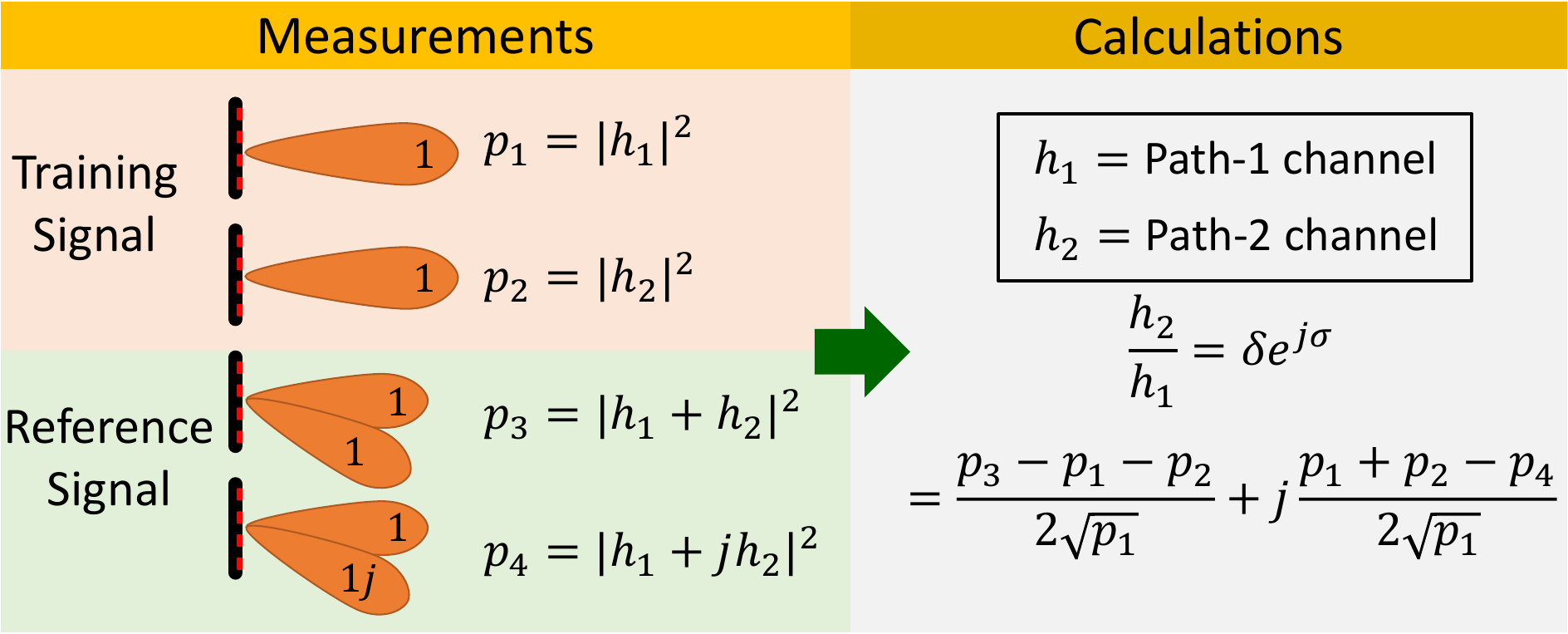}
    \caption{Channel probing procedure to obtain parameters ($\delta, \sigma$) for multi-beam constructive combining that's robust to CFO/SFO impairments.}
    \vspace{-2mm}
    \label{fig:derivation_multibeam}
\end{figure}
First, we observe that only the \textit{relative channel between the two beams} is of interest to us, which means that we can treat $h_1$ as a reference and assume $h_1\in \mathbb{R}^+$ without loss of generality. Then, we observe that the magnitudes $p_1=|h_1|^2$ and $p_2=|h_2|^2$ are already available from the initial beam-training phase. Now, we develop a method to estimate their ratio using just two extra channel probes. Shown in Fig.~\ref{fig:derivation_multibeam}, the gNB uses two RS to probe two 2-beam patterns: first it sets the beam $w(\phi_1,\phi_2,1,0)$, and then $w(\phi_1,\phi_2,1,\frac{\pi}{2})$. For each of the two probes, it estimates the channel magnitudes $p_3$ and $p_4$ as:
\begin{equation}\label{eq:two_probe_cc}
    \begin{split}
        p_3 &= |h_1 + e^{j0} h_2|^2 = |h_1|^2+|h_2|^2+2h_1\operatorname{Re}(h_2)\\
        p_4 &= |h_1  + e^{j\frac{\pi}{2}} h_2|^2 = |h_1|^2+|h_2|^2-2h_1\operatorname{Im}(h_2)
    \end{split}
\end{equation}
where $j=\sqrt{-1}$. Using Eq.~(\ref{eq:two_probe_cc}), we can estimate $\hbf_1$ and $\hbf_2$ individually, and express their ratio as:
\begin{equation}\label{eq:two_probe_result}
    \begin{split}
    \frac{h_2}{h_1} &= \hat{\delta} e^{j\hat{\sigma}} = \frac{p_3-p_1-p_2}{2\sqrt{p_1}}+j\frac{p_1+p_2-p_4}{2\sqrt{p_1}},
    \end{split}
\end{equation}
where $\{\hat{\delta},\hat{\sigma}\}$ are the estimates of the relative amplitude and phase respectively between the two multi-beam paths. Thus, our algorithm estimates the required parameters using only two consecutive RS probes for the two-beam case. The algorithm can be generalized to any $K$ multi-beam by performing two probes for each additional beam and solving Eq.~(\ref{eq:two_probe_result}) to yield $K-1$ relative channels w.r.t $h_1$. In~\cite{palacios2018adaptive}, the authors develop a similar technique, albeit for measuring the per-element channel in a phased array. The channel probing overhead in~\cite{palacios2018adaptive} is proportional to the number of antennas $N$ ($\approx5N$). In contrast, our solution to measure the per-beam channel requires $2(K-1)+K$ probes for $K$ beams (including the $K$ estimates from the beam-training phase) and is independent of the number of antenna elements in the phased array; making it tractable for large arrays.

\noindent $\blacksquare\;$ \textbf{Handling wideband channels:} Typically, the Channel State Information (CSI) is measured across multiple frequency subcarriers over a wide bandwidth. The CSI is readily available for the 5G NR system from the reference signal~\cite{5gnr}. Even for 802.11ad commercial routers, the CSI can be extracted in the firmware~\cite{qualcomm2017wil6210, wang2020demystifying, palacios2018adaptive}. The prior analysis can easily be extended to multiple subcarriers by simply treating each subcarrier's channel independently. We denote $\hbf_1(f)$ and $\hbf_2(f)$ as the wideband CSI across frequency index $f$. We first estimate $\hbf_1(f), \hbf_2(f)$ using Eq.(\ref{eq:two_probe_result}). Then, we formulate an optimization problem that maximizes the average received signal strength overall frequencies to jointly estimate $\{\hat{\delta},\hat{\sigma}\}$ as:
\begin{equation}\label{eq:snroverall}
   \{\hat{\delta}, \hat{\sigma}\} = \argmax_{\delta,\sigma}\;\; \norm{ \hbf_2(f)  - \delta e^{j\sigma} \hbf_1(f)}^2.
\end{equation}

We solve the above to obtain a closed form solution ($\langle\cdot,\cdot\rangle$ is the inner product over the frequency dimension):
\begin{equation}
    \hat{\delta}e^{j \hat{\sigma}} = \langle\hbf_1(f),\hbf_2(f)\rangle/\,\norm{\hbf_1(f)}^2
\end{equation}
which reduces, as expected, to $\hat{\delta}e^{j \hat{\sigma}} = h_2/h_1 $ for a narrowband channel (e.g. in IEEE 802.11ad 60 GHz standard \cite{ieee80211ad}). The optimal phase, amplitude, \& angle will change over time as the user moves. We periodically estimate these parameters using our tracking algorithm discussed in the next section. 




\subsection{Delay phased array beamforming for wideband (channel bandwidth) operation}
Here we explore a new architecture called delay phased array, which uses two phased arrays connected with a network of delay lines and then to a single RF chain. Each phased array creates a single beam in a different direction such that they coherently add to a single receiver. Such architecture can handle multi-path channel bandwidth, providing flat frequency response. 

We first show that a single beam system does not require the new architecture because there is only one path that causes the same delay at each antenna.

\begin{figure}[t!]
    \centering
    \includegraphics[width=0.35\textwidth]{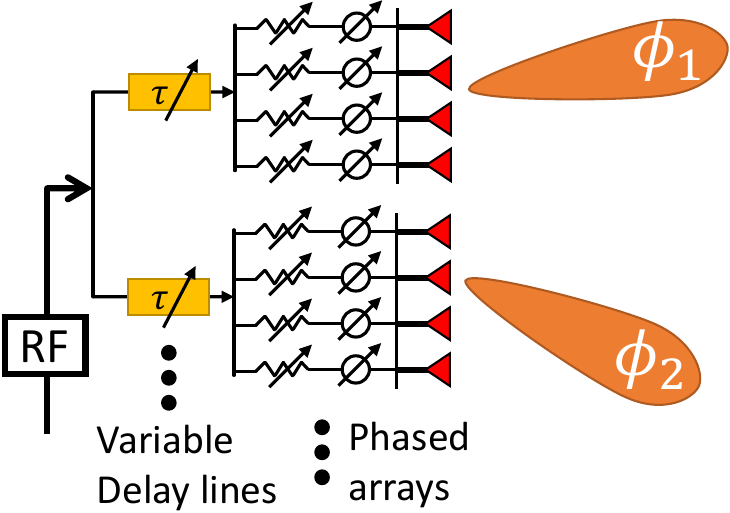}
    \caption{Proposed architecture for delay phased array multi-beamforming to handle wideband transmission. Note that this architecture can solve wideband issue due to multi-path channel. }
    \label{fig:two_arrays_ttd_architecture}
\end{figure}

Let us reconsider the single path channel from $N$ antenna Base station to a single antenna user. Assume the signal direction of departure is $\phi_1$. At the far away location of receiver, the signal travels a different distance from each of the transmit antenna which causes a delay in signal arrival given by $\tau_1$ Considering this delay, we rewrite the expression for single path channel from (\ref{eq:1pathchannel}) to the following expression of $\hbf^{\text{single}}(t,n)$:

\begin{equation}
\begin{split}
        \hbf^{\text{single}}(t,n) 
        &= h_1e^{j2\pi (n-1)\frac{d}{\lambda}\sin(\phi_1)}\delta(t-\tau_1) 
\end{split}
\end{equation}
Here the delay $\tau_1$ is constant for all antennas and can be handled by baseband processing. Therefore, the optimal beamformer for such a single path channel remains a single beam pattern given by (\ref{eq:wphi1}).

\textbf{Multi-beamforming with new delay phased array architecture}
Next, we show that the traditional phased arrays optimal for a single-beam system don't work well for a wideband multi-beam system. Then we propose a new architecture called delay phased array based on delays and phase shifters that achieves optimal multi-beam performance despite a large delay spread of multi-paths. 

The multi-path channel with two paths consists of delays $\tau_1$ and $\tau_2$ as:
\begin{equation}
    h(t,n) = h_1e^{j2\pi (n-1)\frac{d}{\lambda}\sin(\phi_1)}\delta(t-\tau_1) + h_2e^{j2\pi (n-1)\frac{d}{\lambda}\sin(\phi_2)}\delta(t-\tau_2) 
\end{equation}
Here due to different delays of the two paths, the multi-beam beamformer will suffer from wideband issues. That's when two copies of the same signal delayed by a different amount when adds up at the receiver, creates an ineterference pattern at different frequencies. Some frequencies has constructive addition while others may have a destructive interference over a wideband scenario. To solve this wideband issue due to multi-path delays, we propose a new delay phased array architecture shown in Figure \ref{fig:two_arrays_ttd_architecture}. Here we divide the phased array into two or more groups, each for one beam in multibeam. We then create additional delays for each sub-phased-arrays that cancel the difference in delays caused by the channel multi-path. To create a two-beam beamformer in this case, we divide the antennas in two groups of $N/2$ antenna each and connect them with a variable delay line. The expression of two-beam beamformer is:

\begin{equation}
        \wbf_{\phi_1,\phi_2}(n) 
\begin{cases}
    e^{-j2\pi (n-1)\frac{d}{\lambda}\sin(\phi_1)}\delta(t-\Delta\tau_1),& \text{if } n\leq N/2\\
    e^{-j2\pi (n-1)\frac{d}{\lambda}\sin(\phi_2)}\delta(t),              & \text{if } N/2<n\leq N
\end{cases}
\end{equation}
where we apply a delay shift of $\Delta\tau_1$ at the first sub-phased-array to compensate for the multi-path delay difference in two paths.

\begin{figure}[t!]
    \centering
    \includegraphics[width=0.49\textwidth]{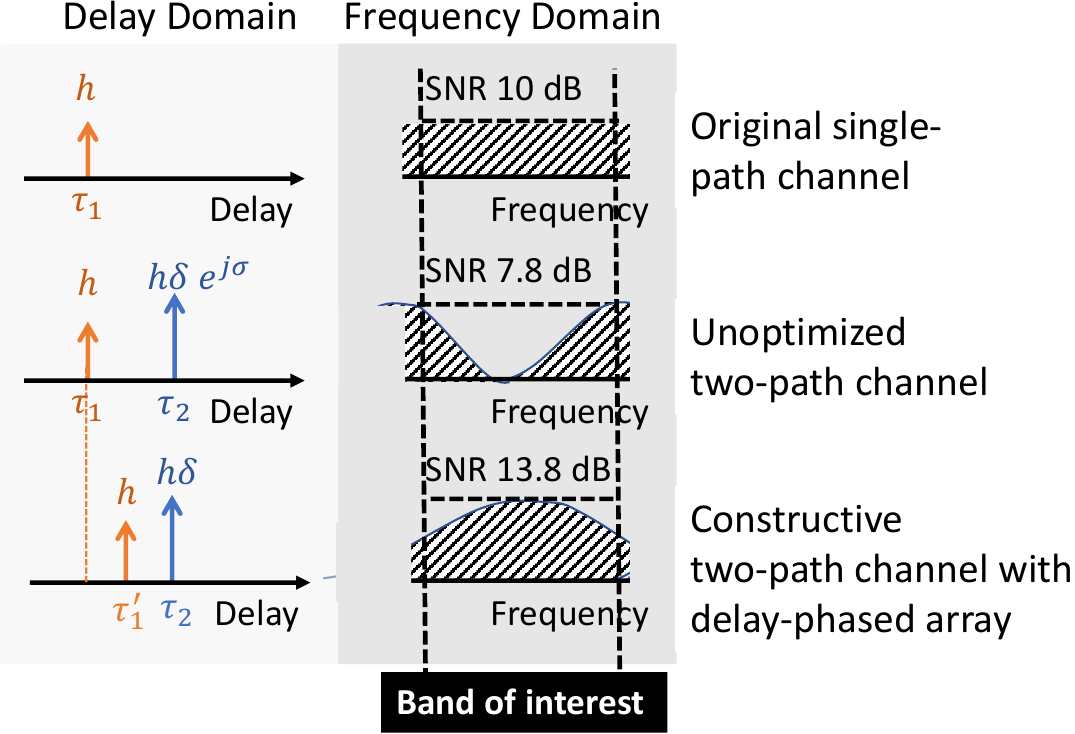}
    \caption{Illustration of the advantages of our delay compensation with delay phased array architecture. We achieve a flat frequency response despite the presence of multi-path delays by compensating for the additional delays caused by the multi-path. }
    \label{fig:delay_compensation_example_freq_domain}
\end{figure}

Such a two-beam beamformer is better than the one in (\ref{eq:beampattern}) because it will work for a wide range of frequencies as shown in Fig. \ref{fig:delay_compensation_example_freq_domain}. Here, we show that a single path channel has a flat frequency response, but a two path channel suffers from imperfect frequency response where some frequency bands are severely attenuated. We also show that by compensating for the multi-path delays, we achieve frequency flat response, while providing 2x gain compared to a single beam system.

\begin{figure}[t!]
    \centering
    \includegraphics[width=0.35\textwidth]{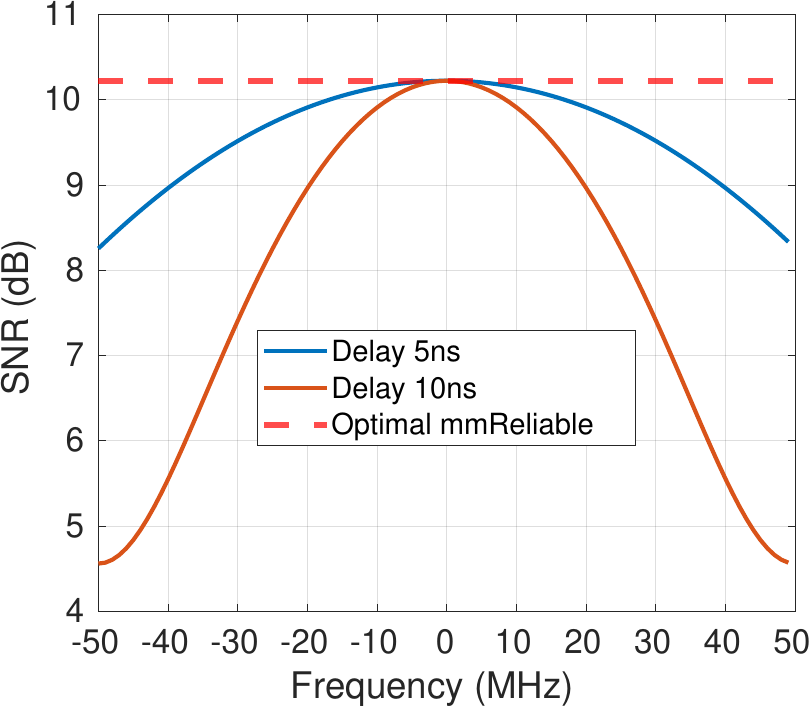}
    \caption{ SNR with range of frequencies. The delay-optimized mmReliable response is flat across all frequency irrespective of any channel delay spread. It also shows for two different delay spread of 5 ns and 10 ns, the SNR suffers at certain frequencies if we do not consider the proposed delay phased array architecture. }
    \label{fig:plot_SNR_with_freq}
\end{figure}

In Fig.~\ref{fig:plot_SNR_with_freq}, we show through simulation that the optimal mmReliable beamformer created using our delay phased array architecture has a flat response across the entire frequency band, while a multi-beam without the proposed delay phased array architecture suffers from low SNR for certain frequencies. We plot two baselines with natural multipath delay spread of 5 ns and 10 ns respectively. While non-optimized \name suffer from signal degradation, delay-optimized \name can compensate for this delay spread and achieve optimal flat performance.

\section{Proactive multi-beam tracking}\label{sec:design}
\begin{figure}[t]
    \centering
    \includegraphics[width=0.48\textwidth]{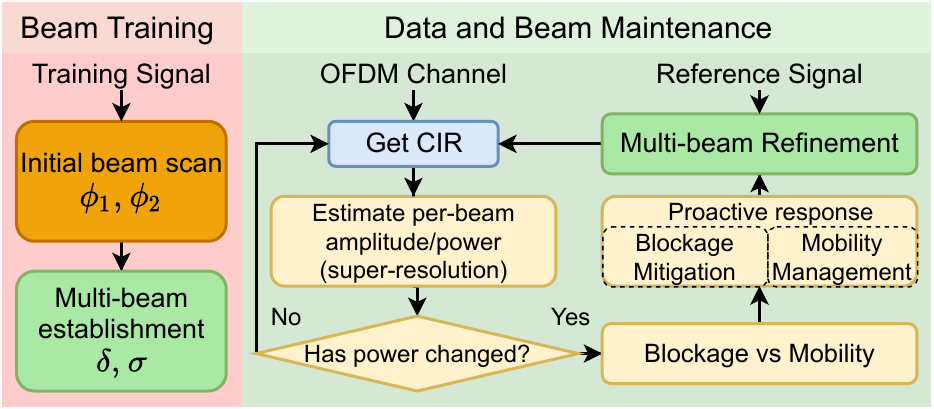}
    \caption{Overview of \name's proactive response against blockage and user mobility. 
    }
    \label{fig:design_overview2} 
\end{figure}

Once a \constructivemultibeam is established, \name maintains it over time to honor the goal of reliable mmWave communication. The purpose of beam maintenance is to keep a high-throughput link even when the user is mobile or sees a random blockage. One choice for beam maintenance could be to periodically repeat the set of procedures in Section \ref{sec:creating_constructive_multibeam} to re-establish \constructivemultibeam with a mobile user. However, it will require repeating the beam-training phase, which incurs significant delays and probing overhead (Section \ref{sec:motivation_beam_maintenance}).
We observe that the effect of blockage and mobility is embedded naturally in the form of variations in the wireless channel.
Thus, our approach is to leverage OFDM channel estimate and reference signaling to maintain a multi-beam link and avoid impending blockage and mobility events while the communication link is active. We show the overall functioning of \name using a flow-chart in Fig.~\ref{fig:design_overview2}.
\name continuously monitors the OFDM channel to identify blockage or mobility on a per-beam basis. User mobility is tracked in the background, and the multi-beam is refined periodically with a low-overhead. When the beam is no longer recoverable by tracking alone (due to accumulated errors in tracking over time), \name re-calibrates the system using the beam training phase.

 \subsection{Proactive Blockage Mitigation}
Mobile blockers interact with the multi-beam system by suddenly occluding one or more active beams. \name analyzes the rate of change of amplitude per beam to detect such events. Empirical evaluation shows that typical blockage events cause per-beam amplitude to degrade by 10 dB in just 10 OFDM symbols. Once a blockage event is detected, \name responds by re-purposing the power on the blocked beam to other unblocked ones. Since there are always multiple beams active, there is no significant impact on link reliability, even if one is blocked.
Said differently, the transceiver reduces the number of beams whenever blockage is detected along some paths. It is rare that all paths are blocked simultaneously; nonetheless, in case of a complete outage, the radio can initiate a new beam training phase to search for other alternate paths or perform a handover~\cite{wei2017pose}. 



\subsection{Proactive User-Mobility Management }
\begin{figure}[!t]
    \centering
    \includegraphics[width=0.99\linewidth]{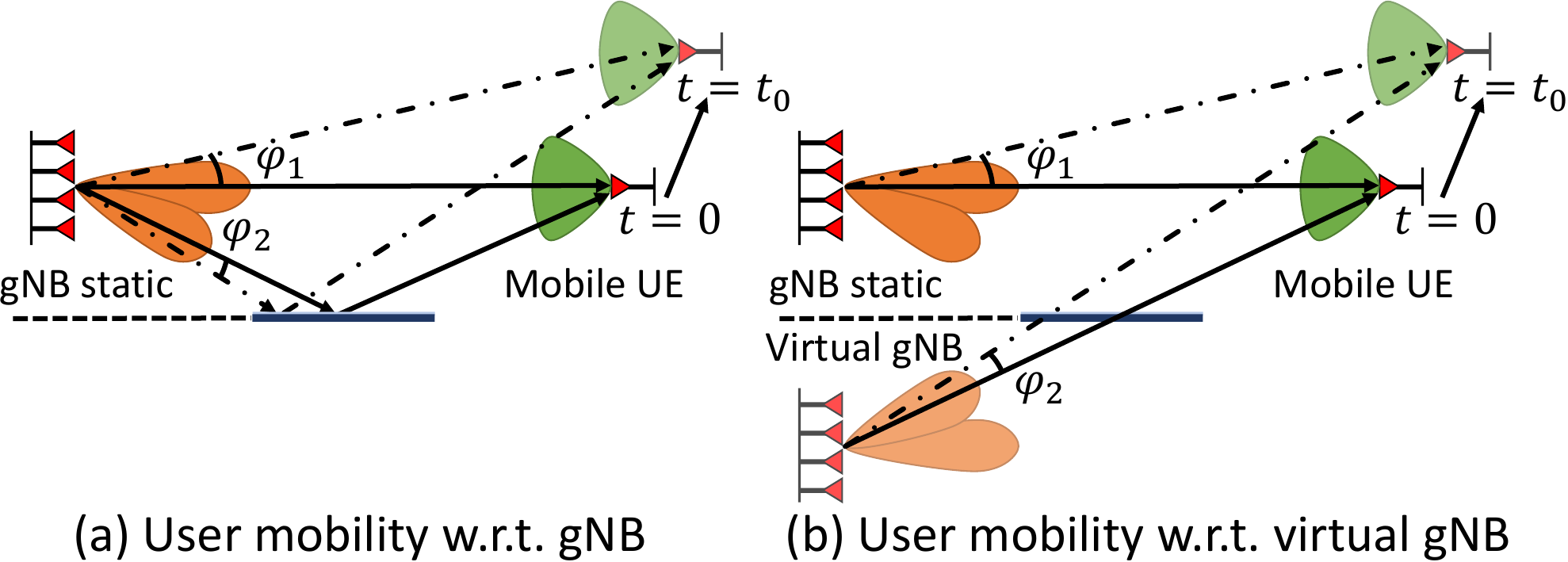}
    \caption{User mobility causes beams in multi-beam misalign with the user. Each beam may undergo a different angular deviation.}
    \label{fig:trans_omniRx}
\end{figure}

When the user is mobile, the initial multi-beam may no longer be supported. A mere angular movement of $14\degree$ would cause a 20dB loss in signal strength leading to outage. A natural solution is to request a new beam training to locate the new user position, but beam training causes high overhead in tedious scanning. 
We propose to proactively track the user mobility using channel measurements and refine the beam periodically. Traditional in-band tracking based on single-beam is not applicable to \name since each beam in multi-beam may undergo different angular deviation making it challenging for tracking as shown in Figure \ref{fig:trans_omniRx}. 

Our insight is that we can track each beam by observing the gradual changes in per-beam power. We propose a model driven approach to estimate per-beam angular deviation from the per-beam power measurements, $P_k(t)$ given by
\begin{equation} 
     P_k(t) = G_T(\phi_k+\varphi_k(t))+G_R+P_T-P_h \; (\text{in dB}).
\end{equation}
where $G_T(\phi_k+\varphi_k(t))$ and $G_R$ are transmit and receive gain respectively, where initial angle $\phi_k$ changes by $\varphi_k(t)$ over time, $P_T$ is transmit power, and $P_h$ is power decay due to channel impairments. We take the difference $P_k(t_0) - P_k(0)$ to find the relative change in the per-beam power as:
\begin{equation}\label{eq:beam_power_difference}
    P_k(t_0) - P_k(0) = G_T(\phi_k+\varphi_k(t_0))-G_T(\phi_k) 
\end{equation}
where we assume that the channel loss $P_h$ (due to path loss or reflection loss) and receiver gain $G_R$ of omni-user is static for the small duration of user motion. Now, to estimate $\varphi_k(t_0)$, our insight is that the direct path power correlates with the beam pattern at the gNB, which is a function of spatial angle:
\begin{equation}
    G_T(\theta) = \frac{\sin(N\theta/2)}{N\sin(\theta)}
\end{equation}
where $N$ is the number of antennas in a uniform linear array and $\theta$ is spatial angle. Therefore, we can use an inverse function to estimate the angle from the measured per-beam power. However, the beam pattern is usually symmetric, and two possible values $\varphi(t_0)$ and $-\varphi(t_0)$ could have caused the observed change in $G_T$. To deal with the ambiguity of the direction of motion, \name tries one of the two possibilities using reference signal probing in the hope that it improves the SNR. If the probe doesn't improve SNR, possibly the other angle is correct, and thus, \name refines the beam to that angle.
The refinement adds an overhead of only one additional probe in addition to $2(K-1)$ probes required for \constructivemultibeam for the K-beam case. \name periodically estimates the value of the per-beam angle, phase, and amplitude and updates the beam pattern at the gNB to realign towards the user.
\review{The tracking algorithm is an important ingredient to support (ii). But the current description is too brief. Please add more information on the delay and AoA estimation and also micro benchmarks on the estimation errors and how they affect the received SNR.}
\newtodo{The reviewer is concerned that the current description of the tracking algorithm is too brief. We will elaborate on our tracking algorithm.}
\subsection{Superresolution: Per-Beam Tracking} \label{sec:establish-multi-beam}
Here we describe how \name estimates the per-beam power required for the tracking algorithm. A simple solution would be to scan single-beams in all $K$ directions and observe the corresponding signal strengths. But, such scanning is intractable due to high overhead in acquiring fine-grained power estimates for efficient tracking. Our idea is to estimate per-beam power using the channel impulse response (CIR) of the current multi-beam in the background without any tedious scanning.  
When receiving a multi-beam transmission, it can be shown that the signal at the receiver consists of a superposition of a delayed and attenuated version of each individual transmit beam. If we consider the delay and attenuation experienced by beam-index $k$ as $\delta_k$ and $\alpha_k$ respectively, then the effective multi-beam channel can be expressed as:
\begin{equation}
\begin{split}
        h_{\text{eff}}& = \sum_k \alpha_k\delta(t-\tau_k),
        \label{eqn:heff}
\end{split}
\end{equation}
where $\alpha_k$ is a function of the transmitter and receiver beam pattern, while $\delta_k$ relates to the Time of Flight (ToF) of each multi-beam channel path. Recall that our goal is to estimate the individual $\alpha_k$ for each beam ($k=1,\ldots, K$). 
\begin{figure} [!t]
    \subfigure{{\includegraphics[width=0.49\columnwidth]{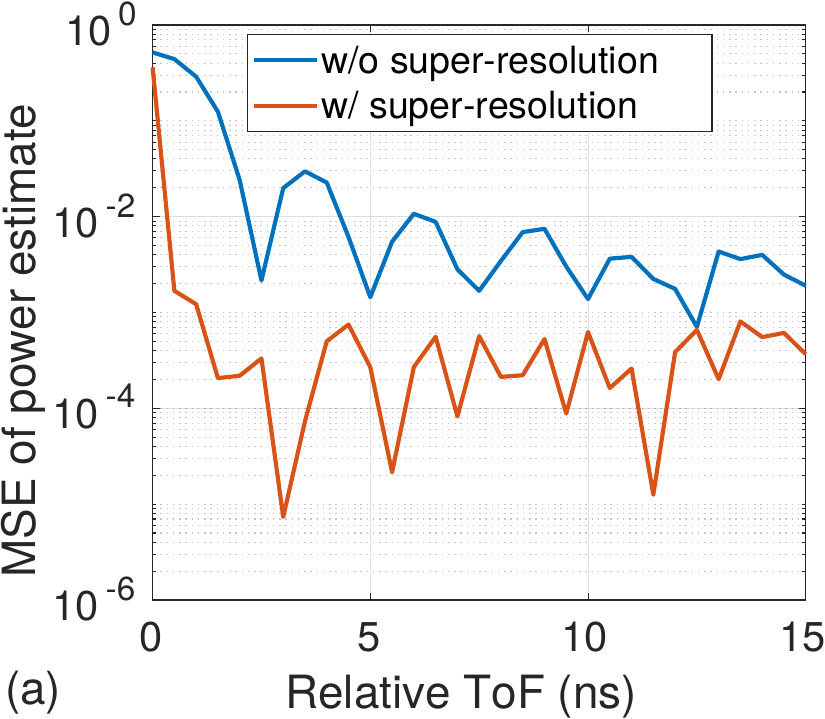}\label{fig:suptimeseries}}}
    \subfigure{{\includegraphics[width=0.49\columnwidth]{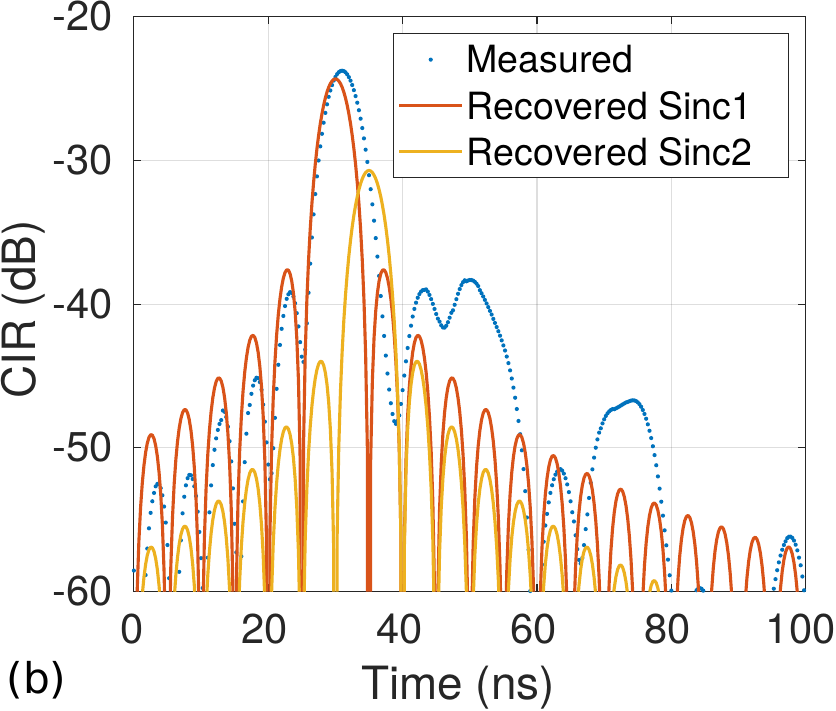}\label{fig:supmse}}}
    \caption{Efficiency of superresolution algorithm: (a) Simulation: Superresolution provides low MSE of per-beam power estimate even when the relative ToF is less than system resolution of 2.5 ns. (b) Hardware measurement: shows two sincs can be recovered from measured combined CIR. }
    \label{fig:superresexample}
 \end{figure}
 
For a frequency-selective wideband system the received signal is sampled at sampling rate and sinc interpolated due to limited bandwidth as follows:
\begin{equation}
\begin{split}
    h_{\text{eff}}[n] &= \sum_k \alpha_k \text{sinc}\left(B(nT_s-\tau_k)\right)\\
\end{split}
\end{equation}
where $B$ is the bandwidth, $T_s$ is the sampling rate of the receiver, and $\alpha_k$ is total signal attenuation along path $k$.  

We can re-write the above formulation as an optimization problem where our objective is to find $\alpha_k$ such that it fits the channel best. Let say the collected CIR is represented by column vector $\mathbf{h}_{\text{CIR}}$.
We solve the following optimization to extract $\alpha = [ \alpha_1 \alpha_2  \ldots \alpha_K]^T$, which amplitude
per beam.
\begin{equation}\label{eq:optim}
    \alpha_{est} = \arg\min_\alpha \|\mathbf{h}_{\text{CIR}}-\textbf{S}\alpha\|^2+\lambda\|\alpha\|^2  
\end{equation}
where the matrix $\textbf{S}$ consists of all the ToF. We are essentially solving a super-resolution problem by fitting a sinc model over the entire CIR response. The optimization is a convex
formulation, with L2 norm regularization. We use standard techniques~\cite{boyd2004convex,cvx} to solve this problem in $100\mu$s. Here, we make a key observation that, after training, the absolute ToF may have changed, but the relative ToF changes slowly. Our key idea is that we shift the $\mathbf{h}_{\text{CIR}}$ first so that the strongest path is shifted to zero delays and since we know relative to the first path, the delay of the second and third path. We can populate the $\textbf{S}$ matrix with only a few columns, thereby achieving accurate and reliable solutions to $\alpha_k$. We account for small variations in relative-ToF by trying few values around the initial value that best fits our model. In summary, we have built an accurate super-resolution algorithm by leveraging the initial relative-ToF information between the multi-beams. 



\noindent $\blacksquare\;$ \textbf{Efficiency of super-resolution algorithm:} 
In Fig. \ref{fig:superresexample}(a), we show that our super-resolution algorithm can indeed achieve high resolution in estimating per-beam amplitude even when the relative ToF is lower than what corresponds to the resolution (2.5 ns for 400 MHz bandwidth). We also measure the channel impulse response through our testbed (6m link with a reflector placed at 30$\degree$) in Fig. \ref{fig:superresexample}(b) and show that our super-resolution can efficiently extract the two sinc which are superimposed in the received CIR.

 \begin{figure}[!t]
    \centering
    \includegraphics[width=\linewidth]{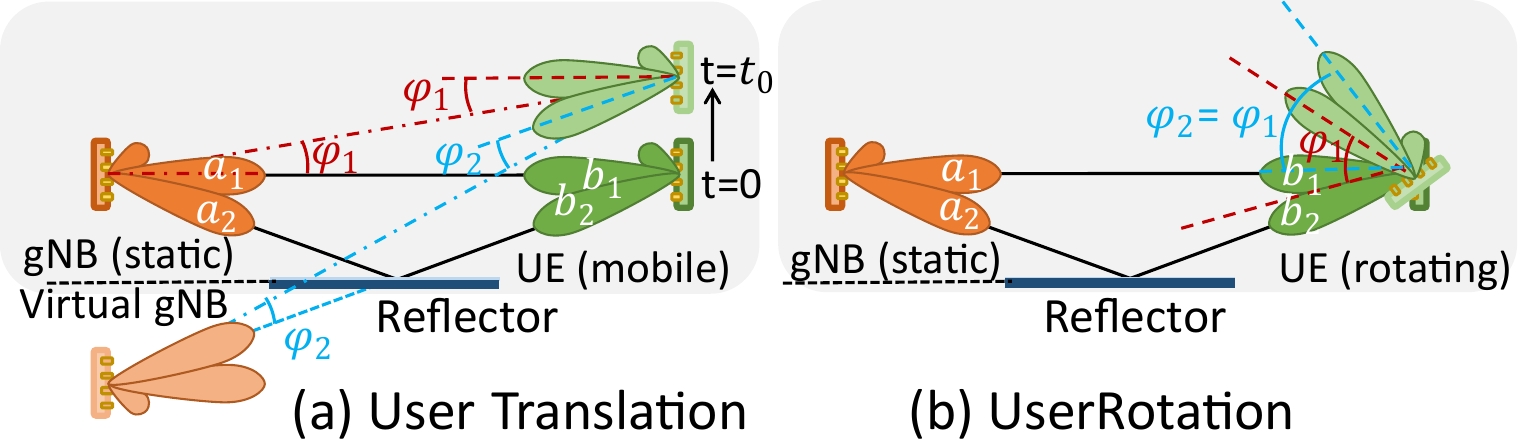}
    \caption{Demonstrating beam misalignment for a multi-beam UE.}
    \label{fig:motion}
\end{figure}
\subsection{Generalization to Multi-beam UE}\label{sec:multibeamUE}

So far we developed \name for a quasi-omni beam pattern at the user equipment (UE). Directional beams may be required at the UE side when the SNR is low, e.g., for longer outdoor links. \name extends naturally to the scenarios where the UEs also have directional beams. In order to do so, \name utilizes the design concepts discussed before, like super-resolution, proactive tracking to mitigate the multi beam link against outages due to mobility, thus addressing the new system level challenges posed by directional UEs. 

The main challenge posed by a directional UE is misalignments on both the UE and gNB side due to UE mobility.  
Consider the scenario in Fig. \ref{fig:motion} (a), in which because of UE moving from t=0 to t=$t_0$, the two beams at the gNB are misaligned by $\varphi_1,\varphi_2$ angle differences respectively. We see a similar misalignment at the UE side as well. If the UE and gNB were able to estimate these misalignment angles, they could compensate for the UE mobility by realigning the beams $a_1,b_1$ and moving them $\varphi_1, -\varphi_1$ respectively, and $a_2,b_2$ will be realigned by $-\varphi_2, \varphi_2$ respectively. Along with estimation of $\varphi_1,\varphi_2$, this step also requires association of beams $a_1,b_1$ and $a_2,b_2$, since there could be fallacious solutions by associating $a_1,b_2$ or $a_2,b_1$ together. Hence, handling UE mobility involves solving the two step problem, by first correctly associating one multi beam to the other, and then estimating the misalignment angles. The first problem follows as a corollary of our super-resolution algorithm, which allows to resolve the per path CIR, and by utilizing the unicity of ToF, we can associate two per path CIRs and hence associate the two beams which are along these respective paths.

In order to mitigate against the misalignments due to user mobility, \name needs to repeatedly estimate and compensate for the misalignment angles. For estimation of the misalignment angles, we develop individual models for tracking translation and rotation similar to our approach in Eq. (\ref{eq:beam_power_difference}). The rotation of UE beam causes changes in UE antenna gain from which we can estimate the angle of rotation using a inverse function (Fig. \ref{fig:motion}). The translation case causes changes in both the gNB's and UE's antenna gain. The received power in this case will be the sum of the two antenna gains. To estimate the angle in this case, we leverage our observation that the beams at gNB and the UE gets misaligned by the same angle (Fig. \ref{fig:motion}). We estimate this angle by inverting the sum of the beam pattern at gNB and UE. In this way, we track the user motion and proactively refine the beams.
\section{Implementation}\label{sec:implementation}
\begin{figure*}[!t]
    \centering
    \includegraphics[width=0.95\textwidth]{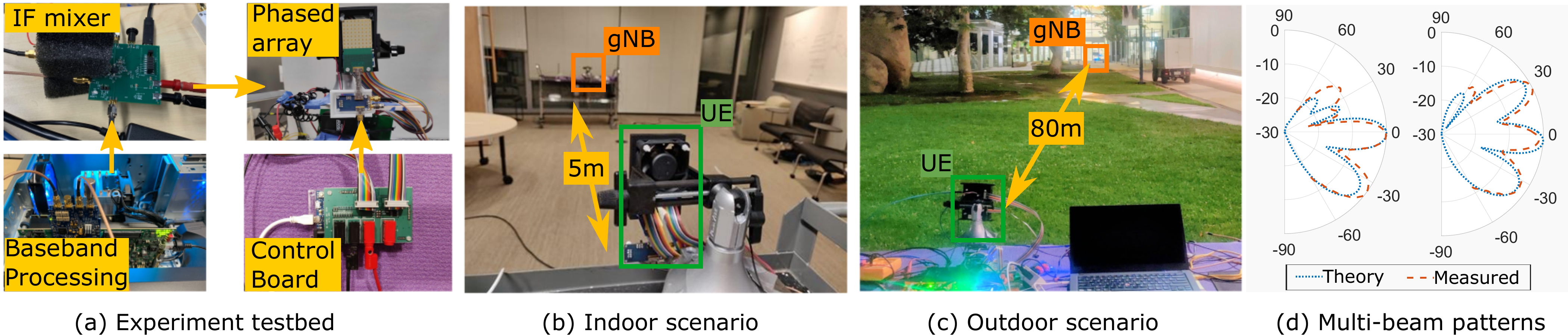}
    \caption{(a) Our experimental testbed with FPGA baseband and 64 element phased array \cite{jain2020mmobile}, (b) indoor 5m and (c) outdoor 80m test scenarios. Our phased array is mounted on a gantry system which provides precise translation and rotation measurements used as ground truth. (d) Comparison of multi-beam patterns from theoretical analysis and our measurements.}
    \label{fig:test_setup}
\end{figure*}

We use a custom 28 GHz mmWave testbed to evaluate the performance of \name. Our testbed consists of two major components as shown in Fig.~\ref{fig:test_setup}(a): (i) an 8$\times$8 mmWave phased array (PA) with real-time beamforming, and (ii) a baseband module that generates and processes standard-compliant OFDM waveforms. For more details on the architectural choices and performance of our testbed, we refer the reader to~\cite{jain2020mmobile}. In this section, we describe components relevant to this work in greater detail.

\subsection{Phased Array and beamforming}
Our 8$\times$8 Phased Array has only one RF chain and is designed in-house~\cite{jain2020mmobile,kibaroglu201864,ma20195g}. It features an integrated up-down converter that operates with an IF frequency of 3.5 GHz. The array is capable of 2 GHz bandwidth centered at 28 GHz and includes amplifiers, image-reject filters in both transmit and receive directions. We use only the azimuth for beamforming and set all the weights along the elevation to the same value. We use a Cinetics Axis-360~\cite{axis360} gantry system to precisely move the array for our evaluations.

\noindent $\blacksquare\;$ \textbf{Beamforming Control:} Inspired by state-of-the-art design~\cite{sadhu2018128}, we use an Artix-7 FPGA~\cite{cmod35t} with custom Verilog modules to program the phased array with high speed (100 us per beam) and accurate timing over an SPI bus. In practical systems, beam-weights that span a limited set of angular directions (typically numbering 64-1024) are programmed into the memory of the FPGA device~\cite{palacios2018adaptive}. Multi-beams require beam-specific amplitude and phase in addition to angular directions, and storing all possible combinations in memory could be intractable. Since multi-beams are linear sums of single-beam-weights, they can be generated using simple addition and multiplication operations on the FPGA. It would allow the fast creation of multi-beam weights while requiring storage of only single-beam weights.

\noindent $\blacksquare\;$ \textbf{Beam Weight Quantization:}
Phase-coherent multi-beam patterns can be generated using minimum 2-bit phase shifters and 1-bit amplitude control (turning antenna on or off)~\cite{orfanidis2002electromagnetic, loch2017zero, palacios2018adaptive} which is available in commercial 802.11ad devices~\cite{qualcomm2017wil6210, talonad7200, madani2021practical} as well as 28 GHz 5G NR~\cite{samsung_array, raychaudhuri2020challenge}. Having higher resolution in phase and amplitude control helps create accurate single-beam and multi-beam patterns that reflect mathematical models. Our array provides 6 bits of phase control and 27 dB of gain control per antenna element similar to commercial products~\cite{samsung_array,raychaudhuri2020challenge,sadhu2018128}\footnote{Verizon, Ericsson, \& other commercial vendors deploy high-resolution 28 GHz phased arrays developed by Samsung, IBM \& others~\cite{verizon_IBM,verizon_Samsung}}. Fig.~\ref{fig:test_setup}(d) shows that our phased arrays generate accurate multi-beam patterns.

\subsection{Baseband module and 5G-NR compliance}Fig.~\ref{fig:test_setup}(d)
Shown in Fig.~\ref{fig:test_setup}(a), we use a 1 GSPs ADC/DAC in conjunction with a KCU-105 FPGA~\cite{kcu,fmcdaq2} for creating 400 MHz baseband OFDM waveforms. We use an off-the-shelf IF mixer~\cite{qorvo} to up/down-convert to the IF frequency of 3.5 GHz. For outdoor experiments (Fig.~\ref{fig:test_setup}(c)), we use a compact setup with a USRP X300 (100 MHz bandwidth)~\cite{jain2020mmobile}. A host PC running MATLAB controls waveform generation, post-processing, and beamforming. We synchronize the array's clock with that of the baseband module to switch beams at intervals with an accuracy of $< 0.1 \mu s$.

\noindent $\blacksquare\;$ \textbf{5G-NR Compliance:} To ensure that \name is compatible with 5G NR standards, we use OFDM waveforms with an effective bandwidth of 400 MHz or 100 MHz and numerology that yields 120 kHz sub-carrier spacing~\cite{3gpp,5gnr}. The beam-training phase in 5G-NR is performed using channel estimates from Synchronization Signal Block (SSB), which repeats with a default period of 20 ms~\citep[TS 38.215]{5GNRRelease16}. \name leverages channel estimates from the Channel State Information Reference Signal (CSI-RS)~\citep[TS 38.215]{5GNRRelease16} for beam-maintenance. CSI-RS can be sent arbitrarily with a spacing ranging from 0.5 ms to 80 ms. A CSI-RS occupies only one symbol (8.93$\mu$s@120 kHz) in the slot and only a few frequency subcarriers~\citep[TS 38.211]{5GNRRelease16}, contributing much less overhead than an SSB. For example, by using only one CSI-RS symbol every 20 ms for beam-maintenance, we incur an overhead of less than 0.04\%. While beam-training cannot be avoided altogether, the maintenance scheme can reduce its periodicity. If the periodicity of a 5 ms long SSB can be extended to 1 second, then the overhead of the entire beam-maintenance process is 0.5\% of channel air time. During post-processing, we use channel estimates only at standard specified intervals to evaluate all algorithms.
\review{he authors should include a more detailed description of their testbed to allow others to reproduce the same results.
}\newtodo{We will provide a more detailed description of our testbed which is elaborated in our previous mmNets’20 paper. }\review{One more issue is that I am not sure if the comparison is really fair. mmReliable uses a flexible phase controlled array and you can tune the phase on each antenna elements to generate arbitrary pattern. But most commerical phased array provides only a limited set of beam patterns. I think the authors might want to make this clear.
}\newtodo{The reviewer asked how multi-beam weights can be realized using a codebook. Since multi-beam weights are a combination of two or more single beam weights, they can be quickly generated in real-time on FPGAs using a codebook of single beams. If computation is at a premium, the multi-beam patterns can be stored in the codebook. We will discuss these practical approaches in the paper. }
\section{Evaluation }\label{sec:results}
We evaluate the performance of \name for various indoor and outdoor settings: a $7m \times 10m$ conference room occupied with wooden furniture, whiteboard, and reflective glass walls (Fig. \ref{fig:test_setup}(b)) and an outdoor $30m-80m$ link next to a large building with glass walls (Fig. \ref{fig:test_setup}(c)).
The Tx module of the testbed is used as a fixed gNB, while the Rx module is movable as a mobile user placed on a cart. 
The phased array antenna of the Tx is kept vertical, facing the user, and parallel to the line of user translation. The precision gantry setup allows 70 cm linear motion and 360-degree rotation about its axis accurate to 1 cm and 0.1$\degree$, respectively, for controlled experiments. The Rx phased array mounted on the gantry (see Fig. \ref{fig:test_setup} (a)) moves as a mobile user with 24$\degree$/s rotation (typical speed for VR headset motion~\cite{qian2018flare}) and $1.5$m/s translation speed in indoor scenarios. For outdoor experiments, we manually move the cart to a predefined trajectory for accurate ground truth information. We also experiment with natural motion for end-end experiments.

We perform multiple experiments of 1-sec duration with mobility and blockage. At the beginning of each experiment, we perform beam training to estimate the angle to the user and establish directional communication. We then run our \constructivemultibeam and user tracking algorithm to realign the beam patterns in real-time as the user moves. Our testbed reports the complex channel per sub-carrier, Signal to Noise Ratio (SNR), and throughput of the wireless link as performance metrics similar to others~\cite{zhou2012mirror, sur201560, sur2016beamspy,zhao2020m,raychaudhuri2020challenge}.

\begin{figure}[!t]
    \centering
    \includegraphics[width=0.4\textwidth]{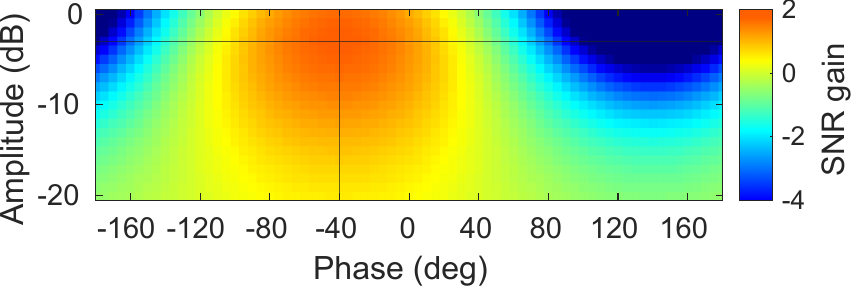}
    \caption{\highlightshep{Sensitivity analysis (simulation):  SNR gain (dB) of 2-beam w.r.t. single-beam for varying 2nd beam's phase $\hat{\sigma}$ and amplitude $\hat{\delta}$ in multi-beam. We show for a -3 dB weaker multipath channel, the multi-beam's highest SNR gain is 1.76 dB which is not very sensitive even for a high phase error of $\pm75\degree$. However, if the phase error is highest ($180\degree$), the SNR drops significantly.}}
    \label{fig:sensitivity}
\end{figure}
 
   \begin{figure*} [!t]
    \subfigure{{\includegraphics[width=0.24\textwidth]{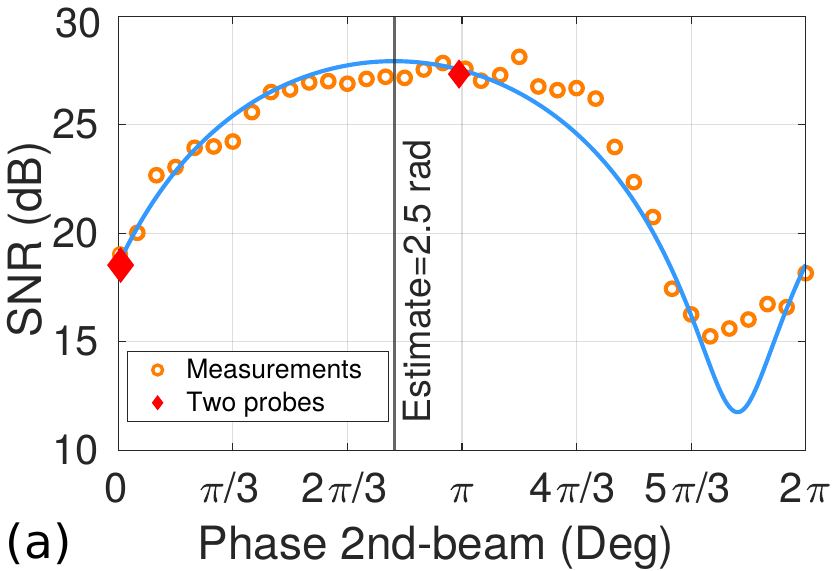}\label{fig:phase_optimization}}} 
    \subfigure{{\includegraphics[width=0.24\textwidth]{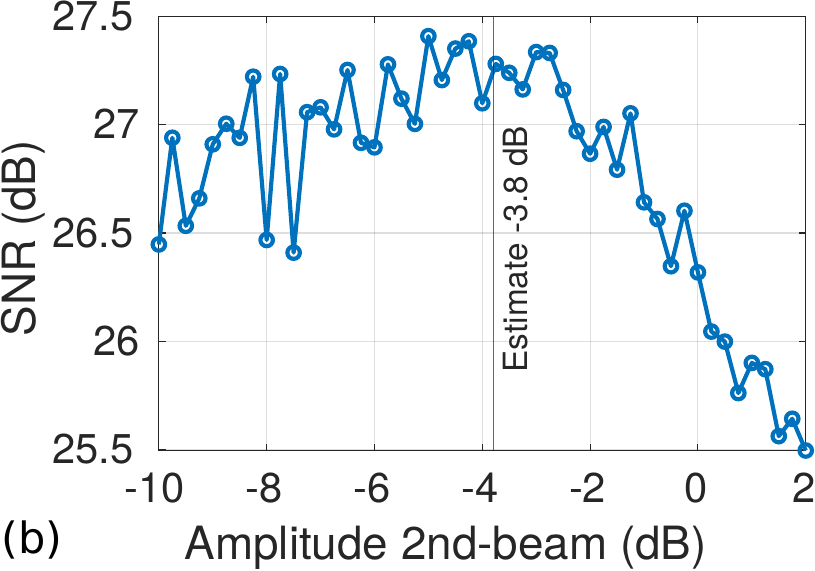}\label{fig:phase_power}}}
    \subfigure{{\includegraphics[width=0.24\textwidth]{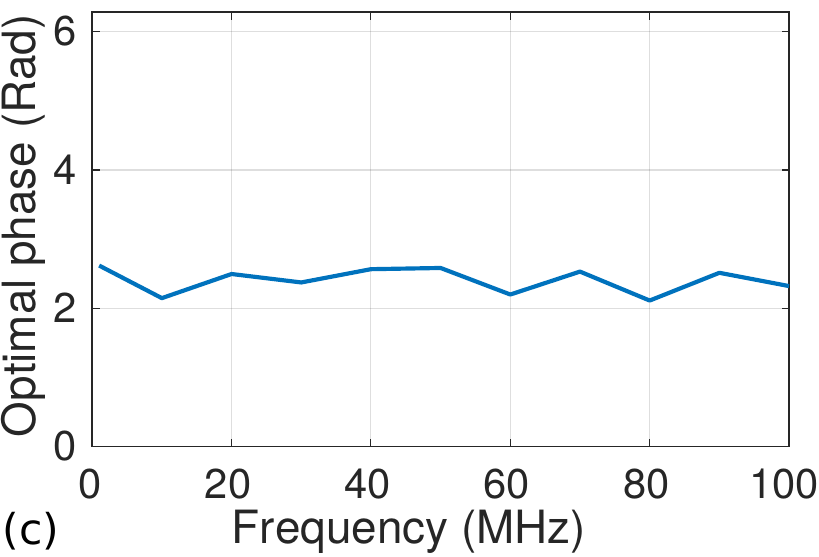}\label{fig:phase_tone}}}
    \subfigure{{\includegraphics[width=0.24\textwidth]{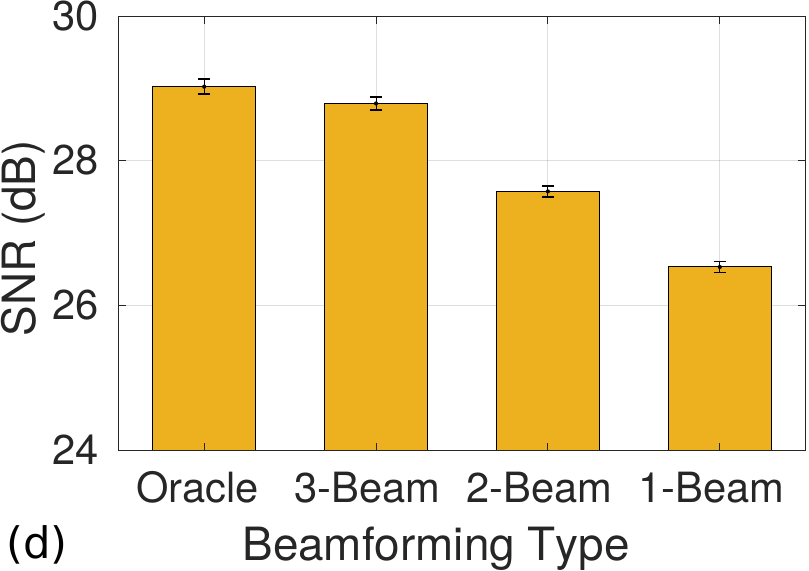}\label{fig:oracle_multibeam}}}
    \caption{(a) SNR measurements with different phases of the 2nd beam in our 2-beam pattern,  (b) SNR measurements with various amplitudes of the 2nd beam through brute-force probing. We also show that our \constructivemultibeam requires only two probes and provides accurate phase and amplitude estimates without brute-force scanning. (c) The optimal phase is stable over a frequency range of 100 MHz. (d) SNR gain of multi-beam against the single beam and oracle (knows the best channel-dependent beam) for a static unblocked link. }
    \label{fig:cophasing}
 \end{figure*}

\subsection{Micro Benchmarks}
Here we present benchmarks for both \constructivemultibeam and proactive user tracking to characterize our system.

\noindent $\blacksquare\;$ \textbf{Sensitivity Analysis of Multi-Beam:} Accurate estimation of phase and amplitude provides the highest SNR; however, it is not very sensitive to estimation errors in these parameters. To quantify the sensitivity, we simulate a two-path channel and set the relative phase and attenuation to $-40\degree$ and $-3$ dB, respectively. We vary the relative phase and amplitude of the 2nd beam in a 2-beam pattern (w.r.t. the first beam) and show the 2D plot of SNR gain in Fig.~\ref{fig:sensitivity}. The highest SNR gain is 1.76 dB for perfect estimation, which reduces gradually with an increasing mismatch between channel parameters and their estimations. The multi-beam SNR exceeds the single-beam SNR and can tolerate errors of $\pm 75\degree$ in phase estimation and amplitude errors of -20 dB.


\begin{figure}
    \centering
    \includegraphics[width=0.35\textwidth]{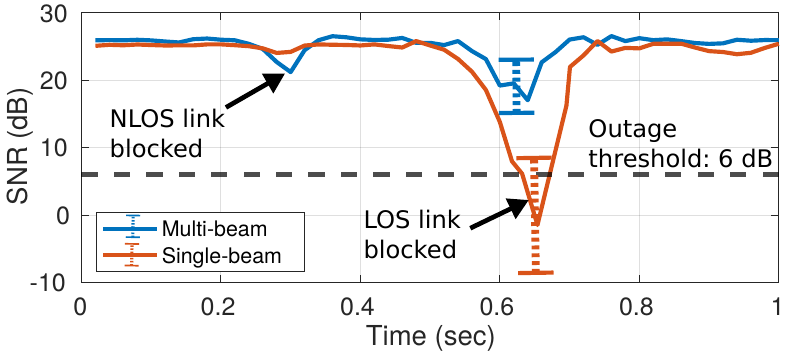}
    \caption{Multi-beam is more reliable and blockage resilient compared to a single beam.}
    \label{fig:snr_blockage_timeseries}
\end{figure}


  \begin{figure*} [!t]
    \subfigure{{\includegraphics[width=0.28\textwidth]{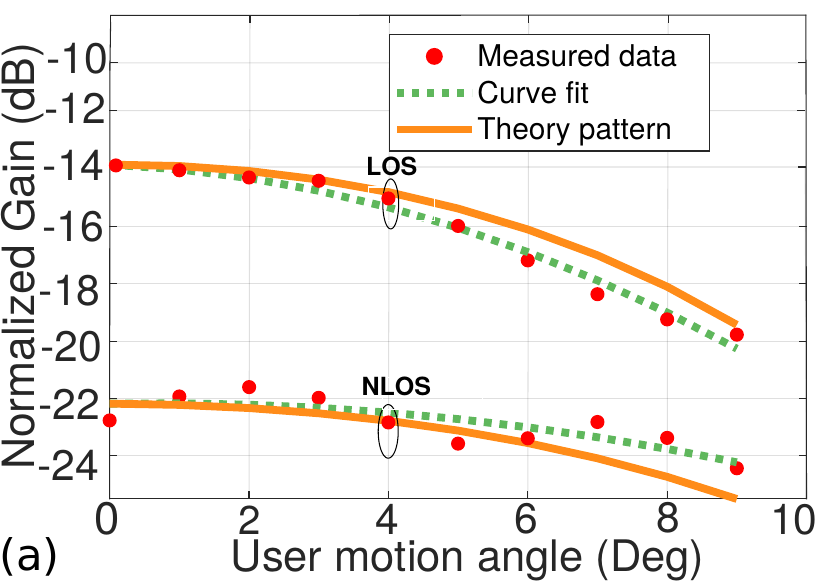}\label{fig:dataRot}}}
    \subfigure{{\includegraphics[width=0.28\textwidth]{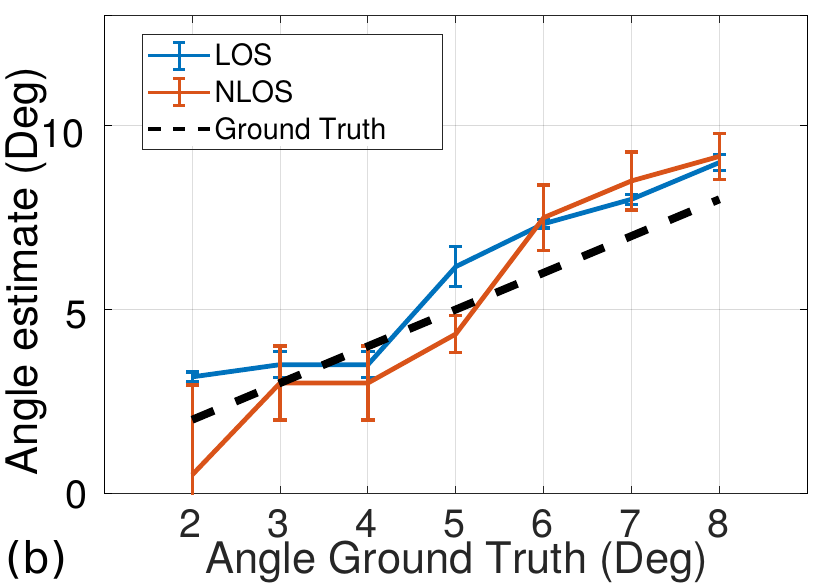}\label{fig:angest1}}}
    \subfigure{{\includegraphics[width=0.28\textwidth]{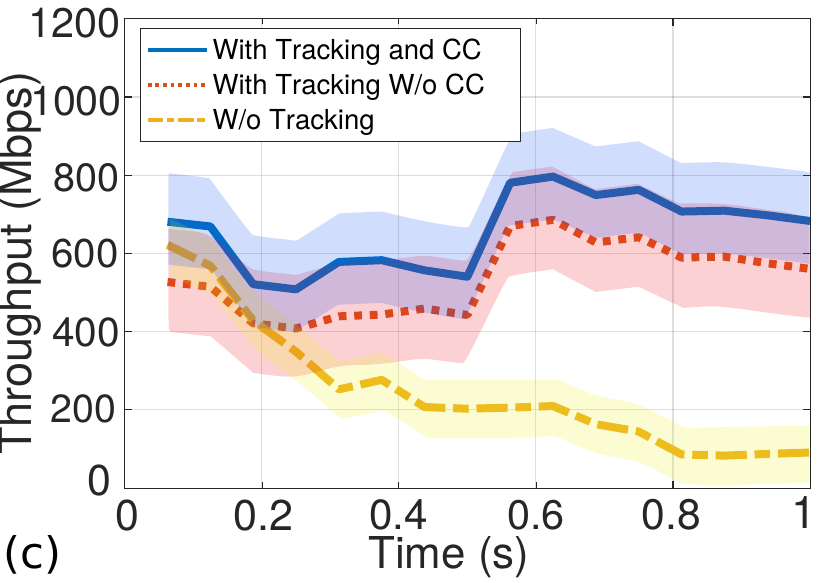}\label{fig:TrackTransTime_avg_decorated}}}
    \caption{(a) Effect of noise and channel impairments on user tracking for UE rotation. We show that even NLOS paths with low signal strength can be tracked effectively. (b) Tracking angle accuracy: \name can estimate the angle within 1$\degree$ error on average. (c) Tracking time series showing the advantage of constructive combining (CC) \& tracking; the shaded region shows standard deviation across multiple experiments, and lines show mean throughput.}
    \label{fig:rottranscase}
 \end{figure*}

\begin{figure*} [!t]
  \subfigure{{\includegraphics[width=0.24\textwidth]{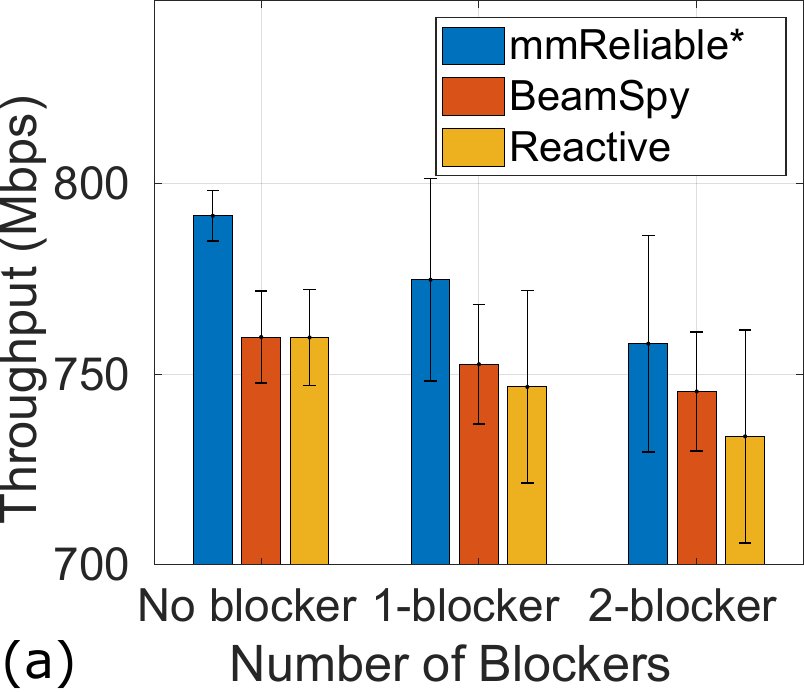}\label{fig:beamspy_compare}}}
    \subfigure{{\includegraphics[width=0.24\textwidth]{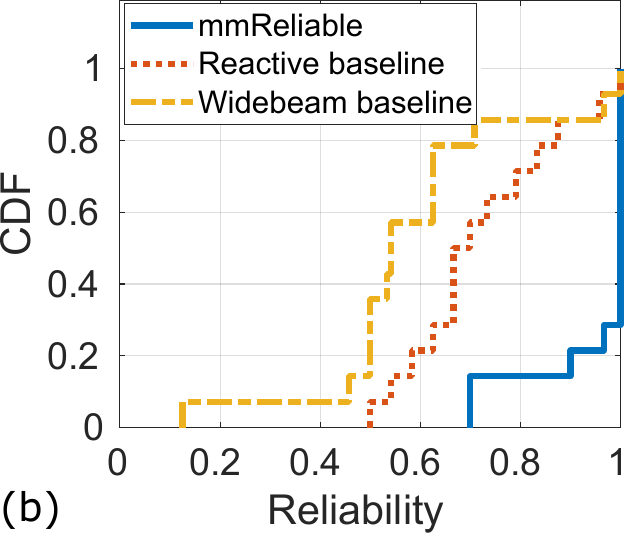}\label{fig:rotbl}}}
    \subfigure{{\includegraphics[width=0.24\textwidth]{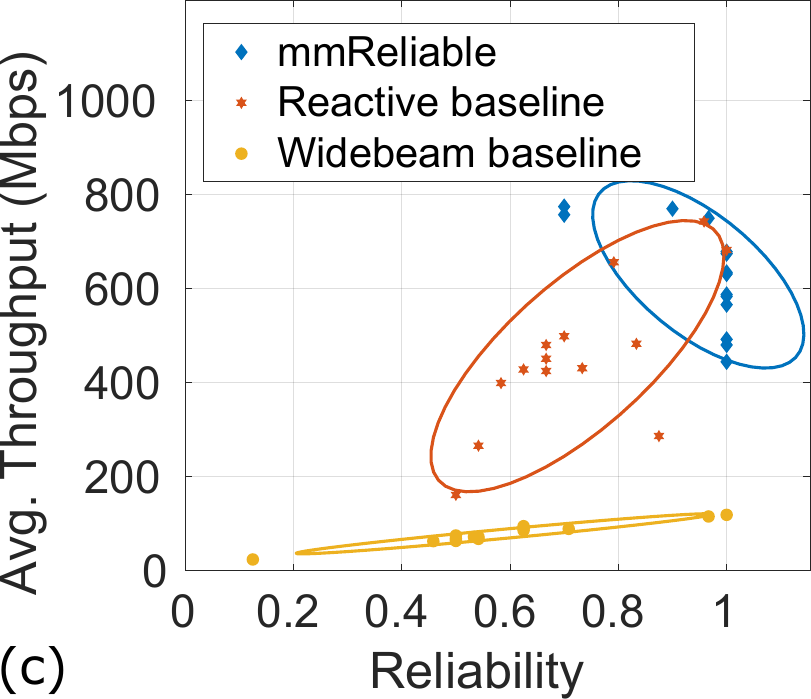}\label{fig:tradeoff}}}
    \subfigure{{\includegraphics[width=0.24\textwidth]{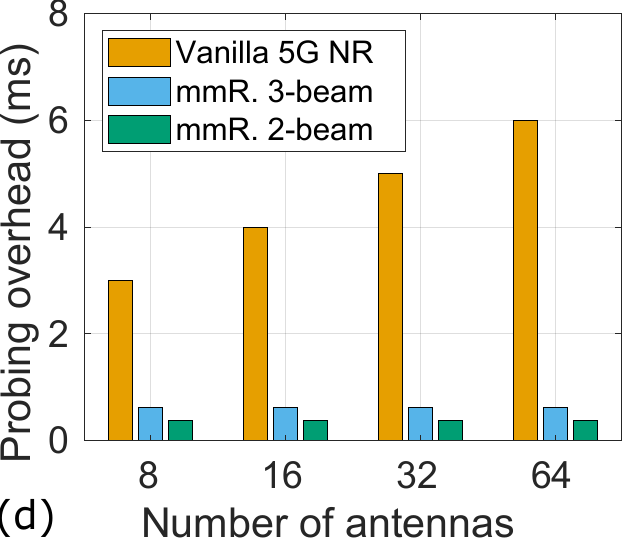}\label{fig:probing_overhead}}}
    \caption{The end-to-end performance gain of \name compared to baselines under different scenarios: (a) Static link with blockages (*evaluates \name without proactive user tracking) (b) Reliability for a mobile link (c) Overall Throughput-Reliability Trade-off; ellipses represent standard deviation in two dimensions. (d) \name (mmR.) has a low probing overhead than vanilla 5G NR.}
    \label{fig:multilobe}
 \end{figure*}

\noindent
$\blacksquare\;$
\textbf{Constructive combining accuracy:} We first show the need for constructive combining in a multi-beam system and then evaluate the performance of \name's algorithm for estimating the phase $\sigma$ and amplitude $\delta$ needed for constructive combining. We establish an indoor multi-beam link with two beams towards a static UE at 7 m along a LOS angle of $0^\circ$ and NLOS angle of $30\degree$, respectively. We set the phase and amplitude of the first beam to be 0 radians, 0 dB, respectively. We then perform an exhaustive scan of the second beam's phase in the range $[0,2\pi]$ and observe the SNR variations in Fig.~\ref{fig:phase_optimization}. We see that the SNR is a maximum of 27 dB with a minor variation of around 1 dB in the range of $\pm 70^\circ$. Any estimate in this range would give a high SNR. But, if the phase estimate is off by $180^\circ$, the SNR degradation could be as significant as 13 dB, which motivates the need for \constructivemultibeam. It also indicates that \name's two-probe method accurately estimates the phase of 2.5 radians, which maximizes the SNR. 
Similarly, we perform an exhaustive scan of the second beam's relative amplitude in the range of [$-10,2$] dB and show the SNR variations in Fig.~\ref{fig:phase_power}. It indicates that the SNR is highest for a range of amplitude around -5 dB to -3 dB. Therefore our estimate of -3.8 dB using the two-probe method is reasonably accurate. Finally, we show the per-beam phase's stability over the 100 MHz frequency range in Fig. \ref{fig:phase_tone}. We observe that phase variation is less than 1 rad, which doesn't impact the SNR gain of our multi-beam system.

\noindent $\blacksquare\;$ \textbf{SNR gain due to \constructivemultibeam:} We now verify that \constructivemultibeam leads to higher SNR than a single beam system, even for a static unblocked link. We reuse our previous set-up and measure that our two-beam multi-beam link provides 1.04 dB SNR improvement on average compared to a single beam oriented towards the LOS path (Fig.~\ref{fig:oracle_multibeam}). The SNR gain is achieved by constructively combining the signal along the two-channel paths. To understand how many beams are required in a multi-beam, we establish an oracle channel-dependent beam that uses the channel at each antenna element of the phased array. Obtaining this oracle beam requires an extensive measurement process with a high overhead~\cite{palacios2018adaptive}. This oracle beam provides an SNR gain of 2.5 dB w.r.t. single beam. We show that our multi-beam pattern with only three beams (utilizing the three most strong channel paths) provides a 2.27 dB gain w.r.t. single beam, which performs similarly to the oracle beam. \textit{This analysis indicates that 3-beams are enough to achieve within 92\% of the optimal beam with significantly less overhead.} 


\noindent $\blacksquare\;$ \textbf{Static link with blockages:} To study the impact of blockage on a multi-beam link, one of the authors walks across our established link multiple times with a consistent walking speed. From Fig.~\ref{fig:snr_blockage_timeseries}, we observe that our multi-beam link with two beams is blocked twice as the blocker passes both the NLOS and LOS link. However, even for the LOS link blockage, the SNR reduction is only 7 dB, which does not cause a link outage. In comparison, a single beam link SNR degrades by 26 dB, below the outage threshold of 6 dB SNR (required for decoding 5G-NR OFDM signals). \textit{Therefore, a multi-beam link is resilient to outages because it avoids a single point of failure.}

\noindent $\blacksquare\;$ \textbf{Accurate per-beam power estimation:} We now test our multi-beam link under mobility. We set up a multi-beam with two beams at the transmitter and rotate the transmit phased array using our gantry. We then collect the channel estimates with our Omni-directional receiver and estimate per-beam power using our super-resolution algorithm. We plot the per-beam power of the two beams in multi-beam with ground truth angle in Fig.~\ref{fig:dataRot}. We observe that per-beam power reduces with the angle of rotation and follows the beam pattern. We observe perturbations in power values due to measurement noise, which may affect tracking performance. To improve tracking accuracy, \name takes time average of power values with a forgetting factor \& fits a quadratic polynomial to smooth the data. With proper scaling, the curve fitting approximates the beam pattern within 1 dB mean error (Fig.~\ref{fig:dataRot}). 


\noindent $\blacksquare\;$ \textbf{Accurate rotation angle estimation:} We next evaluate the performance of our tracking algorithm. We perform multiple experiments with a uniform rotation of the transmitter array. The frequency of rotation is varied over a range of 2$\degree$ to 8$\degree$, and the average angle estimated by our algorithm is reported in Fig.~\ref{fig:rottranscase}(a). The mean angle estimate is accurate to 1$\degree$ for tracking both LOS \& the NLOS link when compared against ground truth.



\noindent $\blacksquare\;$ \textbf{Tracking accuracy time series:} Accurate angle estimation by \name's proactive tracking algorithm helps maintain a high throughput link under user mobility. Figure \ref{fig:TrackTransTime_avg_decorated} shows the throughput of our multi-beam links for a uniform translation of 1 sec duration (at 1.5 m/s speed) under three situations. First, we show that the link throughput degrades from 600 Mbps to $\sim$ 100 Mbps below our outage threshold if no tracking is performed. In contrast, multi-beam tracking with constructive combining (CC) approach maintains a consistently high throughput 1-sec of user mobility. Finally, if we perform tracking alone, without doing constructive combining optimization, the throughput improvements are lower by an average of 100 Mbps compared to the case of tracking plus CC. \textit{Thus, \name provides consistently high throughput due to its super-resolution, per-beam tracking, \& constructive combining algorithms.} 
\subsection{End-to-end Results}
 We present end-to-end results for \name and compare the performance against other baselines \cite{sur2016beamspy, hassanieh2018fast,haider2016mobility}.

\noindent $\blacksquare\;$ \textbf{Multi-beam performance for static link:} 
We first show that \name with \constructivemultibeam and proactive response to blockages can beat other single-beam-based baselines for static links. For this, we implement \name without the user tracking algorithm. We also implement two baselines; (i) BeamSpy~\cite{sur2016beamspy} which detects link blockage by exploiting spatial channel profiles to find the alternate unblocked link without extensive training. (ii) Reactive baseline~\cite{hassanieh2018fast}, which implements a fast beam training in response to the blockage event. Both baselines are based on a single beam, and they suffer from a single point of failure to link blockage and act after an outage is detected. We see from Fig. \ref{fig:beamspy_compare} that \name outperforms both baselines and provides a high throughput mmWave link even under the impact of blockage. \name's throughput drops only by 4\% even when there are two blockers near the beams because of the proactive utilization of reflectors. Since BeamSpy~\cite{sur2016beamspy} was designed for 60 GHz links, we show that \name beats BeamSpy even for 60GHz links in Appendix~\ref{sec:60ghz}.

\noindent $\blacksquare\;$ \textbf{Improvement in Reliability for mobile links:} We perform user translation and rotation experiments in multiple indoor and outdoor environments. As the user moves, a human blocker is introduced midway between gNB and UE, blocking the link for a duration chosen uniformly between 100 ms to 500 ms~\cite{Rappaport2013Millimeter}, over 1-sec experiments. We perform 100 such experiments and combine the results to form one point, and various user mobility and blockage patterns across two indoor environments are reported on the throughput and reliability curve. Fig.~\ref{fig:rotbl} shows that \name achieves close to $100\%$ reliability (median value 1). The reactive baseline suffers from lower reliability of median value 0.65, while the widebeam baseline is at 0.5. Thus, \name achieves its goal of high reliability.

\noindent $\blacksquare\;$ \textbf{Throughput-Reliability Trade-off:} We introduce a new method to evaluate mmWave systems by interpreting the data from the last sub-section using both reliability and throughput metrics. Using both metrics allows us to compare \name with other baselines to assess the \textit{throughput-reliability tradeoff} in Fig.~\ref{fig:tradeoff}. Across all experiments, \name is not significantly affected by throughput loss, while the other benchmarks show degradation due to blockage and user mobility. The results indicate that \name delivers an average throughput improvement of 50\% over the reactive baseline (200 Mbps improvement over 400 Mbps). More importantly, \name offers a consistent throughput with low variations compared to the baselines, essential to many mission-critical applications.  \textit{The evaluation reveals that \name is advantageous over throughput-based design on both throughput \& reliability. }
\review{Figure 14(c) compares the relationships between reliability \& throughput for multiple different schemes. Performing this comparison was a nice addition to the paper. However, the figure and its text are a bit misleading: how are the ovals drawn here? For the alternative schemes, the ovals stop just short of their best values, but for mmReliable, the oval extends well beyond the best points (and excludes the worst points). Moreover, the corresponding text says that "the average reliability is 1 for mmReliable" but for this to be true, all of the reliability values for mmReliable would have to be 1, but that is not the case. Perhaps you meant "median"? In which case, that would appear to bring up the reactive baseline to about 0.7 or 0.75.}
\review{R4: According to 7.1.3. a fixed threshold of 1.5 dB power degradation is used for beam refinement. The overhead vs SNR tradeoff should be discussed. Of course, an aggressive refinement yields a higher average SNR but with the cost of higher overhead!
ToDo: Discuss tradeoff
}

\noindent $\blacksquare\;$ \textbf{Beam probing overhead:} \name reduces the beam probing overhead by maintaining a multi-beam link through tracking and beam refinements. The proposed beam refinement procedure requires three CSI-RS probes for two-beam case: two for estimating $\delta,\sigma$ and one probe for detecting the direction of motion. For the 3-beam case, it only increases to 5 probes. On the other hand, an entire beam scan using SSB in traditional 5G NR requires a large number of SSB probes proportional to the number of spatial directions. One CSI-RS occupies one slot, which is 0.125 ms at 120 kHz sub-carrier spacing, and one SSB takes four slots (0.5 ms). Let's consider the best scanning method, which requires the number of probes proportional to only a logarithmic number of antennas~\cite{hassanieh2018fast}. We show in Fig.~\ref{fig:probing_overhead} that traditional 5G NR suffers from higher probing overhead compared to \name. The 5G NR probing overhead for eight antennas base station is 3 ms, which increases to 6 ms for 64 antennas (because of the high directionality of beam patterns). In contrast, the overhead of \name remains as low as 0.4 ms for 2-beam \& 0.6 ms for 3-beam cases independent of the number of antennas. \textit{Thus, \name is an efficient solution for beam management, which improves mmWave link reliability \& throughput while maintaining low system overhead.}

\section{Related Work}\label{sec:related}

\name is closely related to extensive work in mmWave communication literature, therefore we have grouped the past work into following categories. 

\noindent $\blacksquare\;$ \textbf{Channel-dependent or multi-beamforming:} 
Multi-beam patterns belong to the family of channel-dependent beamforming~\cite{sun2013multi, palacios2018adaptive} where the beamforming weights are obtained to maximize the SNR for a given channel condition. For instance, \cite{sun2013multi} showed through channel-sounding measurements that combining multiple signals across four directions could achieve 28 dB improvement in path-loss at 28 GHz; however, they require multiple phased arrays. Traditionally these optimizations are proposed for a MIMO system that uses channel measurements at each antenna \cite{el2014spatially,heath2016overview,el2012low} for digital beamforming. In contrast, \name focuses on analog beamforming where the signal from each antenna element is connected through a single RF chain. 
ACO~\cite{palacios2018adaptive} proposed a procedure to estimate per-antenna CSI for analog array by designing a specific beam scan procedure.

However, all work in this field incurs high overhead proportional to the number of antennas and cannot be repeated often as the channel changes. In contrast, we propose a multi-beam that approximates the optimal beam pattern and is easy to maintain with low overhead. A similar system \cite{aykin2019multi} proposes \constructivemultibeam for communication that focuses solely on beam-training and multi-beam creation. Specifically, it does not describe any beam-maintenance scheme and therefore would incur high overheads and be unreliable in practice.  
\name is the first system that enables multi-beam constructive combining with low-complexity beam training and continued beam maintenance, mitigating blockage and mobility effects.






\noindent $\blacksquare\;$ \textbf{Blockage  and mobility: }
BeamSpy~\cite{sur2016beamspy} and Beam-forecast \cite{zhou2017beam} provide a model-driven approach to handle blockage and mobility, respectively. However, they suffer from high run-time optimization and table-lookup complexity. To recover from a complete outage, UBig~\cite{sur2018towards} performs efficient handovers, and mmChoir~\cite{zhang2018mmchoir} proposes joint transmission from multiple base stations. These approaches may improve system reliability but add additional signaling overhead. Location estimation using extensive beam training from single or multiple gNBs are proposed in~\cite{haider2016mobility, rasekh2017noncoherent,pefkianakis2018accurate,garcia2020polar,palacios2019leap}. In contrast, \name performs tracking of the user without any training overhead. \cite{loch2017zero} developed a mechanism to detect the direction of user movement and adjust beam direction in small increments. Though they use a 2-lobe pattern to detect mobility, they switch back to a single-beam pattern for communication. The fundamental difference is they use a single beam and are generally reactive. 

\noindent $\blacksquare\;$ \textbf{Beam Training:} A vast literature focus on reducing the initial beam training delay through single-beam \cite{jeong2015random,barati2015directional,zhou2012efficient,zhou2012efficient,barati2016initial, sur201560} or multi-beams \cite{abari2016millimeter,hassanieh2018fast,tsang2011coding, aykin2019smartlink, aykin2019multi}. In contrast, \name focuses on link maintenance and reliability and therefore is complementary to initial beam training approaches. High training overhead may be reduced by taking assistance from device positioning systems \cite{wei2017pose,va2015beam} or light sensors~\cite{haider2018listeer}, out-of-band WiFi~\cite{sur2017wifi,nitsche2015steering}, and multi-user coordination techniques \cite{shao2018two, qiao2014mac}. However, in contrast to \name, they suffer from relatively lower accuracy with added complexity while dependent on external devices. \name high accuracy is achieved based on the super-resolution algorithm combined with the motion tracking algorithm. 



\noindent $\blacksquare\;$ \textbf{Reflecting surfaces:} 
Prior work on active mmWave relay~\cite{abari2017enabling,tan2018enabling} requires additional active devices (antenna array) with beamforming capabilities to relay the signal towards the user. These devices increase cost and power consumption. On the other hand, \name relies entirely on natural reflectors present in the environment. The authors of \cite{zhou2012mirror,genc2010robust,zhao2018improving,khawaja2018coverage, zhu2014demystifying,wei2017facilitating,zhou2019autonomous} leverage reflective surfaces to improve indoor mmWave connectivity \& coverage. Their model is suitable for one-time sensing for initial network deployment, which complements our goal of improving the link reliability in real-time. 
\section{Discussion and Future Work}\label{sec:discussion}
Since multi-beams are a new class of beamforming, designing a robust networking system while multiplexing users is an open challenge. We discuss some limitations of \name and identify possible future research directions.

\noindent \textbf{Strong reflectors:} The functioning of \name depends on the availability of strong reflectors in the environment, otherwise, the multi-beam system falls back to a single-beam system. We envision future deployments where intelligent reflecting surfaces~\cite{dunna2020scattermimo, abari2017enabling, wang2020intelligent,tan2018enabling} are deployed in the environment to engineer strong reflections that improve the throughput and reliability of mmWave links.

\noindent \textbf{Tracking re-calibration:} 
While we can track the angle of movement of both direct and reflected paths using super-resolution and tracking algorithms, the tracking of reflected paths would be affected when the path is blocked, or the reflector vanishes. \name detects such cases by observing per-beam power and reallocates the power along other unaffected beams. If the tracking error accumulates over time, \name needs to initiate a beam training to refresh the system or perform a handover.

\noindent \textbf{Hybrid beamforming and \name:} Using multiple mmWave RF chains and phased arrays can provide degrees of freedom to multiplex users or improve throughput. Prior work~\cite{jog2019many,ghasempour2018multi} has proposed interference-aware spatial multiplexing of beams in different directions for multiple users. These techniques fit with \name in many ways. First, many multi-beams can be created, one from each RF chain, allowing \name's improvements for multiple users. Second, in addition to per-beam phase, amplitude, and angle, multiple RF chains allow for control over group delay, enabling arbitrarily wideband operation. Finally, we can jointly use some spatial beams for enhancing reliability while others for improving multi-user coexistence.

\section{Acknowledgements}\label{sec:acks}

We are grateful to the anonymous reviewers and our shepherd, Prof.
Romit Roy Choudhury, for their insightful feedback. We thank Prof.
Gabriel Rebeiz, UCSD, and his group for the mmWave 5G
phased arrays and transceivers. The research is supported by NSF CCRI \# 1925767.

\noindent



\newpage
\bibliographystyle{unsrt}
\balance
\bibliography{acmart}
\label{lastpage}
\appendix

\section{Constructive Multi-beam is optimal for SNR and throughput for general k-path channel}\label{appendix:A}
Here we show that the multi-beam patterns with appropriate power and phase control are optimal such that they maximize the SNR at the receiver. To demonstrate the multi-beam optimality, we first construct the maximum likelihood (ML) based solution and then show that the multi-beam approaches the ML estimate. Recall the optimal beamforming weights for a general multi-path channel $\hbf$ is given by:






\begin{equation}\label{eq:wmrc2}
    {\wbf}^{\text{opt}}  = \frac{\hbf^*}{||\hbf||},
\end{equation}
which provides the maximum SNR of $\frac{||\hbf||^2P_s }{P_\eta}$. Note that the optimal beamforming weights is dependent on the multi-path channel taps.

Our goal is to show how it compares against the SNR of a single beam and our multi-beam system. We first write the channel in a geometric model representation~\cite{alkhateeb2014channel,sayeed2007maximizing}:
\begin{equation}
    \begin{split}
        \hbf(t) =\sum_{\ell=0}^{L-1} \abf (\phi_\ell) \gamma_\ell e^{2\pi f_c \tau_\ell}\delta(t-\tau_\ell),
    \end{split}
\end{equation}
where we assume the channel is bounded by $L$ paths and $\phi_\ell, \gamma_\ell, \tau_\ell$ represents the AoD, attenuation, and ToF of the $\ell^{th}$ path, respectively. The steering vector $\abf (\phi_\ell)$ represents the relative phase shift introduced by each path at every antenna element (out of $N $ elements) as  $$ \textbf{a} (\phi_\ell) =[1,e^{-j2\pi\frac{d}{\lambda}\sin(\phi_\ell)},\ldots,e^{-j2\pi(N -1)\frac{d}{\lambda}\sin(\phi_\ell)}]^T$$

In the frequency domain, the channel is written at subcarrier $k$ as follow:
\begin{equation}
    \begin{split}
        \hbf(k) = \sum_{\ell=0}^{L-1} \abf (\phi_\ell) \gamma_\ell e^{j 2\pi (f_c +k \Delta f) \tau_\ell} 
    \end{split}
\end{equation}
where $f_c$ is center frequency and $\Delta f$ is the subcarrier spacing. 

The traditional single-beam pattern ignores the multi-path and creates a directional pattern along the strongest path (LOS or one of the reflectors). Assuming $\phi_0$ is the direction of the strongest path, the single-beam antenna weights follows:
\begin{equation}
\begin{split}
     \wbf ^{\text{single}}  &= \abf ^*(\phi_0)/||\abf (\phi_0)|| \\
     &= \frac{1}{\sqrt{N }}[1, e^{j2\pi \frac{d}{\lambda}\sin(\phi_0)},\ldots,  e^{j2\pi(N-1) \frac{d}{\lambda})\sin(\phi_0)}]^T.
\end{split}
\end{equation}

Assuming a directional single-beam pattern severely attenuates other channel paths, the single-beam SNR consists only of the strongest path as:
\begin{equation}
    \text{SNR}^{\text{single}} = \frac{||\gamma_0||^2P_s }{P_\eta}.
\end{equation}
Instead, we design phase-coherent multi-beam patterns which exploits $B$ channel directions out of $L$ paths as follow: 
\begin{equation}
    \wbf ^{\text{multi}} = \frac{\sum_{b=0}^{B-1} \abf ^*(\phi_b) \gamma_b e^{j  \sigma_b}}{||\sum_{b=0}^{B-1} \abf ^*(\phi_b) \gamma_b e^{j  \sigma_b}||},
\end{equation}
where we orient all the $B$ beams in the multi-beam along the $B$ strongest paths in the channel. Note that our multi-beam patterns ensures that per-beam amplitude $\gamma_b$ and phase $ \sigma_b$ (averaged over all frequency subcarriers) is aligned with per-path channel attenuation and phase respectively. We clearly see that when $B=L$, i.e., the multi-beam patterns are oriented along all the paths in a multi-path channel. In this case, the multi-beam weights are same as the optimal weights in (\ref{eq:wmrc2}):
\begin{equation}
     \wbf ^{\text{multi}} = \wbf ^{\text{opt}}, \quad \text{ when } B=L.
\end{equation}

Thus, a multi-beam pattern utilizes $B$ out of $L$ channel paths to improve the SNR as follows:
\begin{equation}
    \text{SNR}^{\text{multi}} = \frac{\sum_{b=0}^{B-1}||\gamma_b||^2P_s}{P_\eta},
\end{equation}
which converges to optimal SNR for $B=L$. We can write the capacity of multibeam-link as
\begin{equation}
    C_{mb} = \log_2(1+\text{SNR}^{\text{multi}})
\end{equation}
The optimal ML beamformer requires per-antenna channel estimate which has a high complexity in terms of beam probing overhead~\cite{palacios2018adaptive}.  \textit{Since mmWave channel is sparse and there are only 1 or 2 strong reflected paths in addition to the direct path, the multi-beam with 2-3 beams provides SNR gain comparable to the optimal beam with significantly lower overhead.}

 \begin{figure}[h]
\subfigure[]{\includegraphics[width=0.21\textwidth]{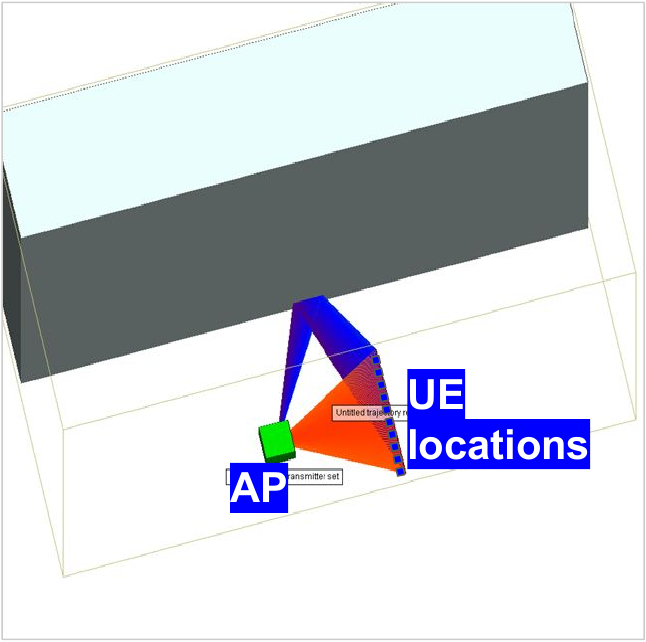}\label{fig:wirelessInsite}}
\subfigure[]{\includegraphics[width=0.25\textwidth]{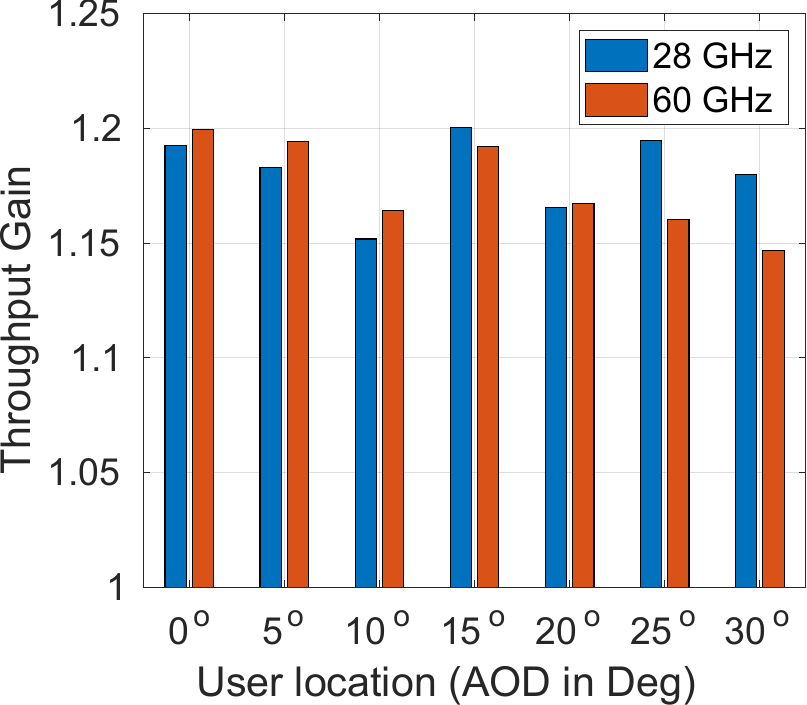}\label{fig:compare28and60}}
\caption{(a) Wireless Insite simulation scenario, (b) Throughput Gain comparing 28 GHz and 60 GHz system for static UE with 10\% blockages.}
\label{fig:wi60ghz}
\end{figure}
\review{The experimental setup is at 28 GHz as opposed to the 60 GHz band. While I understand the hardware limitations of building a large phased array platform at 60 GHz, it is not completely clear whether the results at 28 GHz perfectly translate to 60 GHz since signal propagation characteristics are quite different.}

\section{Comparing \constructivemultibeam performance for 28 GHz and 60 GHz links}\label{sec:60ghz}
\name presents a general framework to improve mmWave throughput and reliability that can also be applied to a 60 GHz link. To compare the two systems, we perform a simulation study using Wireless Insite \cite{wirelessinsite}, as shown in Figure \ref{fig:wi60ghz}. We establish a directional link at 10m with a reflecting surface (made up of concrete material) at $60^o$ (Figure \ref{fig:wirelessInsite}). We set other simulation parameters according to a recent large-scale study of a 60 GHz system~\cite{wang2020demystifying}. The results in Fig.~\ref{fig:compare28and60} show that \name outperforms single-beam based baseline \cite{sur2016beamspy} by 1.18$\times$ gain in throughput for static UE with 10$\%$ blockages. Both 28 GHz and 60 GHz system performs similarly, and the result is consistent across multiple UE locations. Nonetheless, 28 GHz throughput is $4.7\times$ higher than 60 GHz (not shown) for the same bandwidth since 60 GHz links suffer from higher path loss and attenuation due to environmental absorption.
Though 28 GHz performance is better for longer links (AOD $\ge 15^o$) since 60 GHz links suffer from larger path loss and attenuation due to atmospheric absorption.

\end{document}